\documentclass[10,journal]{IEEEtran}
\usepackage{amsmath,amssymb,amsfonts}
\usepackage{algorithmic}
\usepackage{algorithm}
\usepackage{array}
\usepackage[caption=false,font=normalsize,labelfont=sf,textfont=sf]{subfig}
\usepackage{textcomp}
\usepackage{stfloats}
\usepackage{url}
\usepackage{verbatim}
\usepackage{graphicx,psfrag,epsfig}
\usepackage{cite}
\usepackage{amssymb}
\usepackage{amsmath,lipsum}
\usepackage{cuted}
\usepackage{balance}
\usepackage{arydshln}
\usepackage{bm}
\usepackage{amsmath}
\usepackage{amsthm}
\usepackage{subeqnarray}
\usepackage{cases}
\usepackage{xcolor}
\usepackage{soul}
\sethlcolor{green}
\usepackage{xcolor}

\usepackage{graphicx}
\usepackage{tikz}
\usetikzlibrary{arrows.meta}

\graphicspath{{./}}
\newtheorem{Theorem}{Theorem}
\newtheorem{Lemma}{Lemma}

\newtheorem{Definition}{Definition}

\def\linspread{0.93}

\makeatletter
\def\ps@IEEEtitlepagestyle{\ps@empty}
\makeatother

\begin{document}

\title{Movable Antenna Enabled Anti-Jamming: From Local Optimality Guaranteed Design to a Unified Antenna Position Optimization Framework}

%\author{
%	\IEEEauthorblockN{
%		Lebin Chen\IEEEauthorrefmark{1},
%		Ming-Min Zhao\IEEEauthorrefmark{1},
%		Min-Jian Zhao\IEEEauthorrefmark{1},
%		and Rui Zhang\IEEEauthorrefmark{2}
%	}
%	
%	\IEEEauthorblockA{
%		\IEEEauthorrefmark{1}
%		\parbox[t]{0.92\textwidth}{\centering
%			College of Information Science and Electronic Engineering, Zhejiang University, Hangzhou 310027, China;\\
%			Zhejiang Provincial Key Laboratory of Multi-Modal Communication Networks and Intelligent Information Processing, Hangzhou 310027, China;\\
%			National Key Laboratory of Millimeter-Wave and Terahertz Remote Sensing, Zhejiang University, Hangzhou 310027, China\\
%			Email: \{12431101, zmmblack, mjzhao\}@zju.edu.cn
%		}
%	}
%	
%	\IEEEauthorblockA{
%		\IEEEauthorrefmark{2}
%		Department of Electrical and Computer Engineering, National University of Singapore, Singapore 117583\\
%		Email: elezhang@nus.edu.sg
%	}
%}

\author{Lebin Chen, \IEEEmembership{Student Member,~IEEE,} Ming-Min Zhao,
	\IEEEmembership{Senior Member,~IEEE}, 
	Lipeng Zhu,
	\IEEEmembership{Senior Member,~IEEE},
	Zhenyu Xiao,
	\IEEEmembership{Senior Member,~IEEE},
	Min-Jian Zhao, \IEEEmembership{Member,~IEEE},
	and Rui Zhang, \IEEEmembership{Fellow,~IEEE}
	 
	 \thanks{{A preliminary version of this work has been accepted for
	 		presentation at the 2026 IEEE Global Communications Conference \cite{chen2026lowcomplexity}.}}
	 
	\thanks{L. Chen, M. M. Zhao, and M. J. Zhao are with the College of Information Science and Electronic
		Engineering, Zhejiang University, Hangzhou 310027, China,  the National Key Laboratory of Millimeter-Wave and Terahertz Remote Sensing, Hangzhou 310027, China, and also with the Zhejiang Provincial Key Laboratory of Multi‐Modal Communication Networks and Intelligent Information Processing, Hangzhou 310027, China (e-mail:
		12431101@zju.edu.cn; zmmblack@zju.edu.cn; mjzhao@zju.edu.cn).		}
		
				\thanks{Lipeng Zhu is with the State Key Laboratory of CNS/ATM and the School
			of Interdisciplinary Science, Beijing Institute of Technology, Beijing 100081,
			China (e-mail: zhulp@bit.edu.cn).}
		
		\thanks{Zhenyu Xiao is with the School
			of Electronic and Information Engineering and the State Key Laboratory of CNS/ATM, Beihang University, Beijing 100191, China (e-mail: xiaozy@buaa.edu.cn).}

	\thanks{
		R. Zhang is with 
%		School of Science and Engineering, Shenzhen Research Institute of Big Data, The Chinese University of Hong Kong, Shenzhen, Guangdong 518172, China (e-mail: rzhang@cuhk.edu.cn). He is also with 
		the Department of Electrical and Computer Engineering, National University of Singapore, Singapore 117583 (e-mail: elezhang@nus.edu.sg).
%		The work of Rui Zhang is supported in part by The Guangdong Provincial Key Laboratory of Big Data Computing, the National Natural Science Foundation of China (No. 62331022), the 2022 Stable Research Program of Higher Education of China (No. 20220817144726001), and the Guangdong Major Project of Basic and Applied Basic Research (No. 2023B0303000001).
		}
        % <-this % stops a space
%\thanks{}% <-this % stops a space
%\thanks{}
}

% The paper headers
%\markboth{Journal of \LaTeX\ Class Files,~Vol.~14, No.~8, August~2021}%
%{Shell \MakeLowercase{\textit{et al.}}: A Sample Article Using IEEEtran.cls for IEEE Journals}

%\IEEEpubid{0000--0000/00\$00.00~\copyright~2021 IEEE}
% Remember, if you use this you must call \IEEEpubidadjcol in the second
% column for its text to clear the IEEEpubid mark.

\maketitle

\pagestyle{empty}
\thispagestyle{empty}

\begin{abstract}
	Anti-jamming capability is of paramount importance in modern wireless communications, yet conventional receivers are severely constrained by their fixed array geometries, which limit the available spatial degrees of freedom and leave performance highly susceptible to strong interference. By contrast, movable antennas (MAs) provide additional spatial degrees of freedom through antenna-position optimization.
	This paper investigates MA-enabled anti-jamming by jointly optimizing the antenna-position vector (APV) and receive beamformer to maximize the output signal-to-interference-plus-noise ratio (SINR). 
	For the canonical setting with a single jammer and a single line-of-sight (LoS) propagation path on the legitimate link, we develop a damped projected-gradient descent (DPGD) algorithm with low computational complexity, which scales linearly with the number of receive MAs. We establish a generic active strict-saddle landscape for the
	resulting steering-coherence minimization problem and prove that DPGD converges almost surely to a local minimizer rather than merely to a Karush-Kuhn-Tucker (KKT) point. For the general setting with multiple channel paths on the legitimate link and/or multiple jammers, we derive an equivalent ridge least-squares formulation and develop a low-complexity soft-active projected-gradient (SAPG) algorithm, which also scales linearly with the number of receive MAs. 
	We further show that the proposed antenna position optimization framework applies to both minimum variance distortionless response (MVDR) and zero-forcing (ZF) reception. Numerical results validate the proposed algorithms, demonstrating substantial gains in both anti-jamming performance and computational efficiency.
\end{abstract}


\section{Introduction}

\IEEEPARstart{A}{ctive} jamming remains a serious threat to reliable
wireless communications, particularly in dense spectrum-sharing
environments and safety-critical applications such as vehicular
networks, unmanned platforms, and industrial automation
\cite{securitySurvey}. By transmitting intentional interference
during channel training or data transmission, a jammer can corrupt
channel acquisition, reduce the post-combining
signal-to-interference-plus-noise ratio (SINR), and even interrupt
the legitimate link. 
Existing physical-layer anti-jamming approaches mainly rely on jamming-resistant reception and adaptive beamforming \cite{antiJamming,hybridAnti}, often assisted by jammer detection \cite{jammingDetectXL}.
Most existing designs use fixed-position antenna (FPA) arrays, in which only the beamforming weights can be adapted to the interference environment, while the array geometry remains fixed after deployment. Since array geometry directly affects directional discrimination and angular ambiguity \cite{ma2024movableantennaenhancedwireless,fpa1}, a fixed geometry limits the receiver's ability to reshape its spatial response for jamming suppression.

Against this background, movable antenna (MA) systems, also referred to as fluid antenna
systems (FASs), introduce antenna position as an additional spatial
degree of freedom \cite{zhu2024historicalreviewfluidantenna}. By
repositioning the antenna elements or subarrays within a prescribed region, an MA
array can reconfigure its field response without increasing the
number of radio frequency (RF) chains \cite{MAWC,MANF}. Antenna-position optimization has been shown to
improve multiple-input multiple-output (MIMO) capacity, spatial diversity, and multiuser communication
performance over conventional FPAs
\cite{10243545,MAsurvey,MA2,MAHybrid}. The same geometric flexibility has also
been exploited in wireless sensing and integrated sensing and
communication (ISAC) \cite{ISACMA,MAISACStat,FASforISAC1}, 
%\cite{ISACMA,FASforISAC1,R2,R3,MAISACStat,R4}, 
as well as in six-dimensional
MA architectures that jointly control antenna positions and
orientations \cite{6DMA1}. 
%\cite{6DMA1,6DMA2,R5,R6}. 
These studies establish
antenna geometry as a controllable component of the wireless channel.
This capability is particularly relevant to anti-jamming reception,
where repositioning can reduce the spatial coupling between the
desired channel and the jammer subspace before or jointly with
digital receive beamforming.
Exploiting this geometric degree of freedom requires information about the propagation and interference environment. Direction-based anti-jamming receivers have shown that jammer angular information obtained during channel training can be reused for subsequent interference suppression \cite{DirectionBasedJamming,multitarget_MA_CS}. MA-enabled sensing further allows incident directions to be estimated from position-dependent field responses \cite{jammingDetectXL,ma2024movableantennaenhancedwireless}, while recent multi-target sensing studies show that reconfigurable array geometries can distinguish multiple incident directions \cite{MultiTargetSensingInMultipathEnvironments}. Staged channel acquisition and antenna-configuration optimization have also been considered for higher-dimensional MA systems \cite{R6}.

%Exploiting this geometric degree of freedom requires information
%about the propagation and interference environment.
%Direction-based anti-jamming receivers have also demonstrated that
%jammer angular information extracted during channel training can be
%reused to design the subsequent interference-suppression combiner
%\cite{DirectionBasedJamming,multitarget_MA_CS}.
%In this paper, we consider a sensing-assisted framework in which the interference parameters are first estimated and then used for interference suppression.
%Jammer detection can first identify the presence of hostile
%interference, while MA-enabled sensing can provide angular
%information by exploiting position-dependent field responses
%\cite{jammingDetectXL,ma2024movableantennaenhancedwireless}.
%Recent multi-target MA sensing studies further indicate that
%reconfigurable array geometries can support the discrimination of
%multiple incident directions
%\cite{MultiTargetSensingInMultipathEnvironments}. Moreover, staged
%channel acquisition and antenna-configuration optimization have been
%investigated for higher-dimensional MA systems \cite{R6}.
%Motivated by these developments, we assume that the desired-link field-response
%parameters and the jammer directions and received powers are
%estimated before antenna position vector (APV) optimization and remain approximately
%quasi-static during one MA repositioning interval. The resulting
%estimates are then used to jointly optimize the APV and the receive
%beamformer.

Nevertheless, only a few existing studies have directly considered interference
suppression with MA arrays. 
In \cite{ZeroForcing}, the antenna positions and beamforming weights
were jointly designed to achieve full array gain and null steering,
with closed-form APV constructions derived under specific
antenna-number and interference-direction conditions. However, the
analysis focuses on these special configurations without considering
a finite aperture, leaving the general finite aperture
zero-forcing (ZF) based APV optimization problem unaddressed.
%In \cite{TVTantiJamming}, the APV and receive
%beamformer were jointly optimized for multi-jammer SINR maximization
%using a deep-learning-assisted design.
{In \cite{TVTantiJamming}, a special case of the joint APV and
	receive-beamformer SINR maximization problem was considered, where
	the desired signal and each jammer are characterized by a single
	steering direction under finite antenna aperture and minimum-spacing
	constraints. The receive beamformer was obtained analytically from
	a generalized Rayleigh quotient, while the APV was generated from
	the signal directions using a neural network for low-complexity
	online inference.} 
Despite this progress, the algorithmic foundations of MA-enabled
anti-jamming remain relatively underdeveloped. With finite movement
apertures and minimum inter-antenna spacing, APV coupled with beamforming vector optimization becomes
a constrained nonconvex problem, for which achieving both low
computational complexity and reliable solution quality is
nontrivial. 
%The difficulty is further compounded in environments with multiple propagation paths and jammers, where the minimum variance distortionless response (MVDR) receiver couples the desired channel, the interference covariance, and the antenna geometry through position-dependent complex responses.

Motivated by these gaps, we consider an MA-enabled receiver in a general setting allowing multiple propagation paths on the legitimate link and multiple far-field jammers. The receiver jointly optimizes the APV and the minimum variance distortionless response (MVDR) beamformer to maximize the output SINR, subject to the distortionless-response requirement, a finite movement aperture, and minimum inter-antenna spacing.  
The main contributions are summarized as follows:

%Motivated by these gaps, we consider an MA-enabled receiver
%with a general multipath desired channel and multiple far-field
%jammers. The receiver jointly optimizes the APV and the minimum variance distortionless response (MVDR)
%beamformer to maximize the output SINR, subject to a finite movement
%aperture and minimum inter-antenna spacing. 
%Our analysis proceeds from the canonical line-of-sight (LoS) single-jammer case, where the optimization landscape can be characterized explicitly, to the general setting allowing multiple legitimate propagation paths and multiple jammers, where an exact low-dimensional reformulation leads to a unified low-complexity optimization framework.
%The main contributions are summarized as follows:

\begin{itemize}
\item
For the single-jammer case in which the legitimate channel is line-of-sight (LoS), we show that the
MA-enabled MVDR design admits a reduction to a
steering-coherence minimization problem, which reveals how MA
repositioning suppresses jamming through spatial decorrelation.
Based on this structure, we develop a damped projected-gradient
descent (DPGD) algorithm using
bounded isotonic projection, whose
per-iteration complexity scales linearly with the number of receive
MAs. More importantly, we establish a
generic active strict-saddle landscape for the
resulting steering-coherence minimization problem, which shows that non-minimizing critical points are unstable under the proposed DPGD dynamics. This further allows us to prove that under mild conditions, DPGD converges almost surely
to a local minimizer rather than merely to a
Karush-Kuhn-Tucker (KKT) point.
	
\item
For the general setting allowing multipath propagation on the legitimate link and/or multiple jammers, we develop a
unified antenna-position optimization framework with low computational complexity for MVDR and ZF based reception. 
First, an equivalent  
ridge least-squares (LS) reformulation is proposed to reduce the MVDR objective to a
small-scale positive-definite system and enable the APV gradient
evaluation via Danskin's theorem, which avoids large covariance-matrix
inversions and their derivatives. 
Building on this reformulation, we propose a soft-active
projected-gradient (SAPG) algorithm that exploits the geometry of
nearly active APV constraints while maintaining low computational
complexity, which also scales linearly with the number of receive MAs. 
The same LS structure also arises for ZF reception, allowing the
proposed formulation and SAPG algorithm to handle both MVDR and
ZF based APV designs, which enables general finite-aperture APV optimization beyond
the special closed-form null-steering configurations considered in
\cite{ZeroForcing}.

%\item
%Extensive numerical results validate the effectiveness and
%computational efficiency of the proposed algorithms. In the LoS
%single-jammer setting, DPGD achieves nearly the same output SINR as
%the interior-point method (IPM) with substantially lower runtime.
%In the general multipath multi-jammer setting, SAPG consistently
%outperforms the fixed-array, alternating-optimization (AO) \cite{TVTantiJamming}, and
%conventional projected-gradient benchmarks, while approaching the
%performance of IPM at a significantly lower computational cost.

\item
Extensive numerical results support the above analysis and demonstrate the effectiveness of the proposed algorithms. In both the canonical and general settings, the proposed methods achieve output SINR comparable to the interior-point method (IPM) while requiring substantially lower runtime. In the general setting, SAPG further outperforms the fixed-array, alternating-optimization (AO) adopted in \cite{TVTantiJamming}, and conventional projected-gradient benchmarks.

\end{itemize}

The remainder of this paper is organized as follows.
Section~II presents the system model and the optimization problem. Section~III studies the LoS
single-jammer case, while Section~IV considers the general multipath multi-jammer case. Section~V provides numerical results, and
conclusions are drawn in Section VI.

%The remainder of this paper is organized as follows.
%Section~II presents the system model and formulates the joint MVDR
%beamformer and APV optimization problem. Section~III studies the LoS
%single-jammer case, derives the steering-coherence formulation,
%establishes its active strict-saddle landscape, and develops the DPGD
%algorithm with an almost-sure local-optimality guarantee.
%Section~IV considers the general multipath multi-jammer case,
%derives the ridge-residual gradient oracle, presents the SAPG
%algorithm, and establishes the connection between the MVDR and ZF
%formulations. Section~V provides numerical results, and
%Section~VI concludes the paper.

\textit{Notations:}
Scalars, vectors, and matrices are denoted by italic lower-case
letters, bold-face lower-case letters, and bold-face upper-case
letters, respectively. 
The symbol $j$ denotes the imaginary unit, i.e., $j^2=-1$.
The operators $(\cdot)^T$, $(\cdot)^*$, $(\cdot)^H$,
$(\cdot)^{-1}$, and $(\cdot)^\dagger$ denote the transpose,
complex conjugate, conjugate transpose, matrix inverse, and
Moore--Penrose pseudoinverse, respectively.
The inner product of two complex vectors is
$\langle\mathbf a,\mathbf b\rangle\triangleq\mathbf a^H\mathbf b$.
The operators $\Re\{\cdot\}$ and $\Im\{\cdot\}$ return the real and
imaginary parts, respectively.
The notation $\|\cdot\|_2$ denotes the Euclidean norm for vectors and
the spectral norm for matrices, while $|\cdot|$ denotes either the
magnitude of a scalar or the cardinality of a set, as determined by
the context.
The symbols $\mathbf I_N$, $\mathbf 0$, and $\mathbf 1$ denote the
$N\times N$ identity matrix, an all-zero vector or matrix, and an
all-one vector of appropriate dimensions, respectively.
The sets of real and complex vectors of dimension $N$ are denoted by
$\mathbb R^N$ and $\mathbb C^N$, respectively.
The expectation operator is denoted by $\mathbb E\{\cdot\}$, and
$\mathcal{CN}(0,\sigma^2)$ denotes a zero-mean circularly symmetric
complex Gaussian distribution with variance $\sigma^2$.
The operators $\operatorname{rank}(\cdot)$ and $\operatorname{diag}(\cdot)$
denote the matrix rank and diagonal-matrix operator,
respectively. For Hermitian matrices, $\mathbf A\succeq\mathbf B$
means that $\mathbf A-\mathbf B$ is positive semidefinite.
For a closed convex set $\mathcal C$, $\Pi_{\mathcal C}(\mathbf x)$,
$\delta_{\mathcal C}(\mathbf x)$, and
$N_{\mathcal C}(\mathbf x)$ denote the Euclidean projection of $\mathbf x$ onto $\mathcal C$,
the indicator function of $\mathcal C$, and the normal cone to $\mathcal C$ at $\mathbf x$, respectively.
The limiting subdifferential of a function $f$ is denoted by
$\partial f$. For a smooth manifold $\mathcal M$,
$T_{\mathcal M}(\mathbf x)$ and
$\nabla_{\mathcal M}^{2}f(\mathbf x)$ denote its tangent space and
the Hessian of the restriction of $f$ to $\mathcal M$, respectively.
For a function $f:\mathcal X\to\mathbb R$ and a subset
$\mathcal M\subseteq\mathcal X$, $f|_{\mathcal M}:\mathcal M\to
\mathbb R$ denotes the restriction of $f$ to $\mathcal M$, i.e.,
$f|_{\mathcal M}(\mathbf x)=f(\mathbf x)$ for all
$\mathbf x\in\mathcal M$.

\begin{figure}[t]
	\centering
	\includegraphics[width=3in]{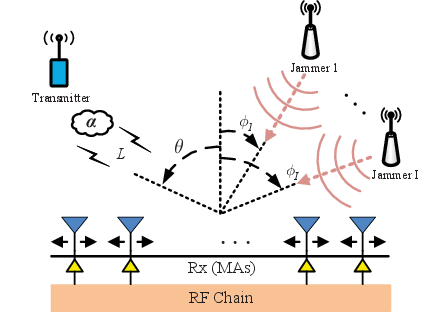}
	\caption{Illustration of MA-enabled receiver under jamming attacks.}
	\label{model}
	\vspace{-0cm}
\end{figure}

\section{System Model and Problem Formulation}

\subsection{System Model}

We consider an MA-enabled anti-jamming communication system consisting of a  legitimate transmitter, $I$ far-field jammers, and an MA receiver equipped with $N_r$ MAs. The channels are assumed to be narrow-band and quasi-static, so that the MA repositioning overhead can be tolerated within the channel coherence time. Let $\mathbf{x}=[x_1,x_2,\ldots,x_{N_r}]^T\in\mathbb{R}^{N_r}$ denote the APV of the receive MA array. 
For an impinging plane wave with azimuth angle $\theta$, the array steering vector is given by
\begin{equation}
	\label{eq:a_theta_nlos}
	\mathbf{a}(\mathbf{x},\theta)
	=
	\begin{bmatrix}
		e^{jkx_1\sin\theta},
		e^{jkx_2\sin\theta},
		\ldots,
		e^{jkx_{N_r}\sin\theta}
	\end{bmatrix}^{T},
\end{equation}
where $k=2\pi/\lambda$ denotes the wavenumber and $\lambda$ is the carrier wavelength.

We adopt a field-response-based channel model~\cite{MAWC,10243545}. The legitimate channel contains $L$ propagation paths. The azimuth angle of the $\ell$-th path is denoted by $\theta_\ell$, and the corresponding complex path gain is denoted by $\alpha_\ell$. Let $\boldsymbol{\alpha}=[\alpha_1,\alpha_2,\ldots,\alpha_L]^T\in\mathbb{C}^{L\times 1}$ and
\begin{equation}
	\label{eq:A_x_def_nlos}
	\mathbf{A}(\mathbf{x})
	\triangleq
	[
	\mathbf{a}(\mathbf{x},\theta_1),
	\mathbf{a}(\mathbf{x},\theta_2),
	\ldots,
	\mathbf{a}(\mathbf{x},\theta_L)
	].
\end{equation}
Then the desired channel vector between the legitimate transmitter and the MA receiver is
\begin{equation}
	\label{eq:h0_nlos}
	\mathbf{h}_0(\mathbf{x})
	=
	\mathbf{A}(\mathbf{x})\boldsymbol{\alpha}.
\end{equation}
The desired channel is assumed to be known at the receiver through channel estimation~\cite{hybridAnti,10243545}.

The received signals collected by the MAs are processed by a digital beamformer $\mathbf{w}\in\mathbb{C}^{N_r\times1}$. The beamformer output at time $t$ is expressed as
\begin{equation}
	\label{eq:y_out_nlos}
	y(\mathbf{x},t)
	=
	\mathbf{w}^{H}\mathbf{r}_y(\mathbf{x},t),
\end{equation}
where the received signal vector is
\begin{equation}
	\label{eq:r_t_nlos}
	\mathbf{r}_y(\mathbf{x},t)
	=
	\mathbf{h}_0(\mathbf{x})s_0(t)
	+
	\sum_{i=1}^{I}\mathbf{g}_i(\mathbf{x})s_i(t)
	+
	\mathbf{n}(t).
\end{equation}
Here, $\mathbf{g}_i(\mathbf{x})\in\mathbb{C}^{N_r\times1}$ is the channel vector from the $i$-th jammer to the receiver, and $\mathbf{n}(t)\in\mathbb{C}^{N_r\times1}$ denotes the noise vector whose elements follow $\mathcal{CN}(0,\sigma_n^2)$.

Over each quasi-static interval, the desired symbol process $\{s_0(t)\}$ is modeled as zero-mean and wide-sense stationary (WSS) with average power $\sigma_s^{2}\triangleq\mathbb{E}\{|s_0(t)|^{2}\}$. The $i$-th jammer symbol process $\{s_i(t)\}$ is also WSS with average power $\sigma_i^{2}\triangleq\mathbb{E}\{|s_i(t)|^{2}\}$, $i=1,\ldots,I$. The desired signal, the jamming signals, and the noise are mutually independent.

The jammers are assumed to impinge on the receiver from possibly different angles. The channel vector of the $i$-th jammer is modeled as~\cite{hybridAnti}
\begin{equation}
	\label{eq:gi_known}
	\mathbf{g}_i(\mathbf{x})
	=
	\zeta_i\mathbf{a}(\mathbf{x},\phi_i),
\end{equation}
where $\zeta_i$ denotes the complex jamming channel gain and $\phi_i$ denotes the angle of arrival (AoA) of the $i$-th jammer. 
We assume that the jammer AoAs and received powers can be estimated before antenna repositioning through jammer sensing and parameter estimation \cite{DirectionBasedJamming,multitarget_MA_CS,MultiTargetSensingInMultipathEnvironments}. These estimates are assumed to remain approximately unchanged during one MA repositioning interval.

The output SINR is
\begin{equation}
	\label{eq:sinr_nlos}
	\begin{aligned}
		\mathrm{SINR}
		&=
		\frac{
			\mathbb{E}\left\{
			\left|
			\mathbf{w}^{H}\mathbf{h}_0(\mathbf{x})s_0(t)
			\right|^{2}
			\right\}
		}
		{
			\mathbb{E}\left\{
			\left|
			\mathbf{w}^{H}
			\left(
			\sum_{i=1}^{I}\mathbf{g}_i(\mathbf{x})s_i(t)
			+
			\mathbf{n}(t)
			\right)
			\right|^{2}
			\right\}
		}\\
		&=
		\frac{
			\sigma_s^{2}
			\left|
			\mathbf{w}^{H}\mathbf{h}_0(\mathbf{x})
			\right|^{2}
		}
		{
			\mathbf{w}^{H}
			\mathbf{R}_{i+n}(\mathbf{x})
			\mathbf{w}
		},
	\end{aligned}
\end{equation}
where
\begin{equation}
	\label{eq:R_in_nlos}
	\begin{aligned}
		\mathbf{R}_{i+n}(\mathbf{x})
		&=
		\mathbb{E}
		\left\{
		\mathbf{p}(\mathbf{x},t)
		\mathbf{p}^{H}(\mathbf{x},t)
		\right\}\\
		&=
		\sum_{i=1}^{I}
		\sigma_i^{2}
		\mathbf{g}_i(\mathbf{x})\mathbf{g}_i^{H}(\mathbf{x})
		+
		\sigma_n^{2}\mathbf{I}_{N_r}
	\end{aligned}
\end{equation}
is the interference-plus-noise covariance matrix, with
\begin{equation}
	\label{eq:p_xt_def_nlos}
	\mathbf{p}(\mathbf{x},t)
	\triangleq
	\sum_{i=1}^{I}\mathbf{g}_i(\mathbf{x})s_i(t)
	+
	\mathbf{n}(t).
\end{equation}

\subsection{Problem Formulation}

We consider a joint receive beamformer and APV optimization problem under the MVDR framework~\cite{AdaptiveArrays}. The design objective is to minimize the output interference-plus-noise power while imposing a distortionless response toward the desired channel and satisfying the MA geometry constraints, i.e.,
\begin{equation}
	\label{eq:8}
	\begin{aligned}
		\min_{\mathbf{w},\mathbf{x}}\quad
		& \mathbf{w}^{H}\mathbf{R}_{i+n}(\mathbf{x})\mathbf{w}\\
		\text{s.t.}\quad
		& \left|\mathbf{w}^{H}\mathbf{h}_0(\mathbf{x})\right|=1,\\
		& |x_m-x_n|\ge d,\quad 0\le x_i\le D_x,\\
		& 1\le m,n,i\le N_r,\quad m\neq n.
	\end{aligned}
\end{equation}
The first constraint enforces a unit-gain distortionless response for the desired signal.\footnote{Since the legitimate transmitter is equipped with a single antenna, the desired link carries a single spatial stream. In this
setting, the linear minimum mean-square error (LMMSE) combiner is collinear with the MVDR beamformer. Specifically, the LMMSE combiner is proportional to $(\sigma_s^2\mathbf{h}_0(\mathbf{x})\mathbf{h}_0^H(\mathbf{x})+\mathbf{R}_{i+n}(\mathbf{x}))^{-1}\mathbf{h}_0(\mathbf{x})$, which is further proportional to $\mathbf{R}_{i+n}^{-1}(\mathbf{x})\mathbf{h}_0(\mathbf{x})$ by the matrix inversion lemma. Since the output SINR is invariant to any nonzero complex scaling of the receive vector, the MVDR and LMMSE combiners achieve the same output SINR. We use the MVDR normalization because it explicitly imposes a unit response toward the desired channel.}
The parameter $d$ is the minimum inter-MA spacing, typically chosen as $d=\lambda/2$ to reduce mutual coupling, and $D_x$ is the available aperture. We assume $D_x>(N_r-1)d$, so the feasible set has a nonempty relative interior beyond the minimum-spacing configuration.

Due to antenna interchangeability, the ordering $x_1<x_2<\cdots<x_{N_r}$ is imposed without loss of optimality. Under this ordering, the geometry constraints can be equivalently written as the following linear inequalities:
\begin{equation}
	\label{eq:Xset_nlos}
	\mathcal{X}
	\triangleq
	\big\{
	\mathbf{x}\in\mathbb{R}^{N_r}:
	\mathbf{U}\mathbf{x}\preceq\mathbf{l}
	\big\},
\end{equation}
where
\begin{equation}
	\label{eq:U_nlos}
	\mathbf{U}
	=
	\left[\begin{array}{cccccc}
		1 & -1 & 0 & \cdots & 0 & 0\\
		0 & 1 & -1 & \cdots & 0 & 0\\
		\vdots & \vdots & \vdots & \ddots & \vdots & \vdots\\
		0 & 0 & 0 & \cdots & 1 & -1\\
		-1 & 0 & 0 & \cdots & 0 & 0\\
		0 & 0 & 0 & \cdots & 0 & 1
	\end{array}\right]_{(N_r+1)\times N_r},
\end{equation}
and
\begin{equation}
	\label{eq:l_nlos}
	\mathbf{l}
	=
	\left[
	-d,\ -d,\ \ldots,\ -d,\ 0,\ D_x
	\right]^T.
\end{equation}

For any fixed APV $\mathbf{x}$, the optimal MVDR beamformer has the following closed-form solution~\cite{AdaptiveArrays}:
\begin{equation}
	\label{eq:wopt_nlos}
	\mathbf{w}_{\mathrm{opt}}(\mathbf{x})
	=
	\frac{
		\mathbf{R}_{i+n}^{-1}(\mathbf{x})\mathbf{h}_0(\mathbf{x})
	}
	{
		\mathbf{h}_0^{H}(\mathbf{x})
		\mathbf{R}_{i+n}^{-1}(\mathbf{x})
		\mathbf{h}_0(\mathbf{x})
	}.
\end{equation}
Substituting \eqref{eq:wopt_nlos} into \eqref{eq:8} eliminates $\mathbf{w}$ and yields the APV-only problem as given by
\begin{equation}
	\label{eq:P0_nlos}
	\max_{\mathbf{x}\in\mathcal{X}}\quad
	g(\mathbf{x})
	\triangleq
	\mathbf{h}_0^{H}(\mathbf{x})
	\mathbf{R}_{i+n}^{-1}(\mathbf{x})
	\mathbf{h}_0(\mathbf{x}).
\end{equation}
Problem \eqref{eq:P0_nlos} is highly nonconvex because both
$\mathbf h_0(\mathbf x)$ and $\mathbf R_{i+n}(\mathbf x)$ depend
nonlinearly on the MA positions through complex steering responses.
Existing studies have derived closed-form APV constructions only under
specific antenna-number and interference-direction conditions
\cite{ZeroForcing}, or employed deep-learning-assisted iterative joint
optimization \cite{TVTantiJamming}. To address the above difficulties, Section~III first considers the
setting with a single LoS propagation path
on the legitimate link and a single jammer, and the DPGD algorithm is proposed with an almost-sure
local-optimality guarantee. 
Section~IV then addresses the general setting with multiple
legitimate propagation paths and/or multiple jammers through the proposed SAPG algorithm, forming
a unified antenna-position optimization framework applicable to both
MVDR and ZF reception.

\section{LoS Channel and Single Jammer Case}
\label{LoSsection}

Following the above roadmap, we first consider the canonical case with one desired LoS path and one jammer. Despite its simpler channel structure, the APV optimization remains challenging because the antenna positions are coupled through a highly nonconvex phasor objective and must simultaneously satisfy the minimum-spacing and aperture constraints. In this case, the MVDR objective can nevertheless be reduced exactly to steering-coherence minimization, which reveals a more tractable structure. Moreover, a common shift of all antenna positions leaves the objective unchanged, and we therefore fix one antenna as a reference to remove this unnecessary degree of freedom. Since a first-order method may still converge to a stationary point that is not locally optimal, we characterize the non-minimizing stationary points of the reduced problem and develop a DPGD algorithm with a pool adjacent violators algorithm (PAVA) based projection. The proposed algorithm has linear per-iteration complexity and converges almost surely to a local minimizer rather than merely to a KKT point. The active-face analysis in this case also motivates the slack-aware design used later for the general case.

\subsection{MVDR Objective in Terms of Steering Coherence}
We first specialize the general model in \eqref{eq:h0_nlos} and \eqref{eq:R_in_nlos} to the case with one desired LoS path and one jammer, i.e., $L=1$ and $I=1$. The desired and jamming channels are then given by $\mathbf{h}_0(\mathbf{x})=\alpha_1\mathbf{a}(\mathbf{x},\theta_1)$ and $\mathbf{g}_1(\mathbf{x})=\zeta_1\mathbf{a}(\mathbf{x},\phi_1)$, respectively. Substituting these channel expressions into the MVDR gain in \eqref{eq:P0_nlos} makes its dependence on the APV $\mathbf{x}$ explicit, and the interference-plus-noise covariance reduces to
\begin{equation}
	\label{eq:special_R_compact}
	\mathbf{R}_{i+n}(\mathbf{x})
	=
	\sigma_n^2
	\left(
	\mathbf{I}_{N_r}
	+
	\gamma
	\mathbf{a}(\mathbf{x},\phi_1)
	\mathbf{a}^{H}(\mathbf{x},\phi_1)
	\right),
\end{equation}
where $
	\gamma
	\triangleq
	{\sigma_1^2|\zeta_1|^2}/{\sigma_n^2}.
$
Since \eqref{eq:special_R_compact} is a rank-one perturbation of the identity and $\|\mathbf{a}(\mathbf{x},\theta)\|_2^2=N_r$, the Sherman-Morrison identity gives
\begin{equation}
	\label{eq:special_Rinv_simplified}
	\mathbf{R}_{i+n}^{-1}(\mathbf{x})
	=
	\frac{1}{\sigma_n^2}
	\left(
	\mathbf{I}_{N_r}
	-
	\frac{\gamma}{1+\gamma N_r}
	\mathbf{a}(\mathbf{x},\phi_1)\mathbf{a}^{H}(\mathbf{x},\phi_1)
	\right).
\end{equation}
Substituting \eqref{eq:special_Rinv_simplified} and $\mathbf{h}_0(\mathbf{x})=\alpha_1\mathbf{a}(\mathbf{x},\theta_1)$  into the objective of \eqref{eq:P0_nlos} yields
\begin{equation}
	\label{eq:special_g_star_closed_form}
	g(\mathbf{x})
	=
	\frac{|\alpha_1|^2}{\sigma_n^2}
	\left(
	N_r
	-
	\frac{\gamma}{1+\gamma N_r}
	\left|
	\mathbf{a}^{H}(\mathbf{x},\phi_1)
	\mathbf{a}(\mathbf{x},\theta_1)
	\right|^2
	\right).
\end{equation}
We further define
\begin{equation}
	\label{eq:special_rho_def}
	\rho(\mathbf{x})
	\triangleq
	\mathbf{a}^{H}(\mathbf{x},\phi_1)
	\mathbf{a}(\mathbf{x},\theta_1)
	=
	\sum_{n=1}^{N_r}
	e^{j a x_n},
\end{equation}
where
$
	a
	\triangleq
	k(\sin\theta_1-\sin\phi_1)
$ is the phase slope.
Thus, maximizing the MVDR gain in \eqref{eq:special_g_star_closed_form} is equivalent to the following steering-coherence minimization problem:
\begin{equation}
	\label{eq:special_equiv_x_problem}
	\min_{\mathbf{x}\in\mathcal{X}}\quad
	f_a(\mathbf{x})
	\triangleq
	\left|
	\sum_{n=1}^{N_r}
	e^{j a x_n}
	\right|^2.
\end{equation}
This equivalence reveals the main anti-jamming role of MA repositioning in the LoS single-jammer setting, i.e., it decorrelates the desired steering vector from the jammer steering vector. 
If $a=0$, i.e., $\sin\theta_1=\sin\phi_1$, the legitimate signal
and the jammer arrive from the same spatial direction as observed
by the receive array. 
They are therefore spatially
indistinguishable to the receive array, so MA repositioning cannot
provide spatial interference suppression under the considered
far-field model.\footnote{In the near field, two sources arriving from the same angle can still produce different array responses if they are located at different ranges.  Extending the present analysis to this near-field setting is an interesting direction for future work.  } 
Accordingly, the objective in
\eqref{eq:special_equiv_x_problem} is a constant, and thus we focus on
the meaningful case $a\ne0$.

The expression in \eqref{eq:special_g_star_closed_form} also clarifies how APV optimization affects the final MVDR output. Since \eqref{eq:wopt_nlos} gives the minimum output interference-plus-noise power under the distortionless constraint, the resulting output SINR is
\begin{equation}
	\label{eq:mvdr_sinr_closed_form}
	\mathrm{SINR}(\mathbf{x})
	=
	\sigma_s^2 g(\mathbf{x})
	=
	\Gamma_s
	\left(
	N_r
	-
	\frac{\gamma}{1+\gamma N_r}f_a(\mathbf{x})
	\right),
\end{equation}
where $\Gamma_s\triangleq\sigma_s^2|\alpha_1|^2/\sigma_n^2$ denotes the desired link signal-to-noise ratio (SNR) factor and $\gamma=\sigma_1^2|\zeta_1|^2/\sigma_n^2$ denotes the jammer-to-noise ratio (JNR). Hence, the desired link SNR only scales the MVDR output SINR, while the APV-dependent loss is completely determined by the steering coherence term $f_a(\mathbf{x})$. 
Since $0\le f_a(\mathbf{x})\le N_r^2$, the MVDR output SINR satisfies 
\begin{equation} 
	\label{eq:mvdr_sinr_bounds} 
	\frac{\Gamma_s N_r}{1+\gamma N_r} \le \mathrm{SINR}(\mathbf{x}) \le \Gamma_s N_r.
\end{equation} 
The upper bound corresponds to perfect steering orthogonality, while the lower bound corresponds to the case where the desired and jamming steering vectors are fully coherent.

This expression also shows why reducing $f_a(\mathbf{x})$ is especially important when the jammer power is large. For a fixed APV, differentiating \eqref{eq:mvdr_sinr_closed_form} with respect to the JNR $\gamma$ gives
\begin{equation}
	\label{eq:sinr_jnr_sensitivity}
	\frac{\partial\,\mathrm{SINR}(\mathbf{x})}{\partial\gamma}
	=
	-\frac{\Gamma_s f_a(\mathbf{x})}{(1+\gamma N_r)^2}
	\le 0.
\end{equation}
%Thus, an APV with smaller steering coherence suffers less SINR degradation as the JNR $\gamma$ increases. This is the main role of MA repositioning in the considered MVDR receiver, i.e., it does not alter the nominal array-gain term $N_r$, but reduces the jammer-induced coherence penalty in \eqref{eq:mvdr_sinr_closed_form}.
Hence, a smaller $f_a(\mathbf x)$ renders the output SINR less sensitive to an increase in the JNR $\gamma$.

\subsection{Landscape Structure of the Reduced APV Optimization Problem}

The above analysis shows that the APV design is equivalent to minimizing the steering coherence penalty $f_a(\mathbf{x})$ in \eqref{eq:special_equiv_x_problem}. 
{Several special cases of this problem admit closed-form solutions. For example, when $N_r\le3$, the small number of optimization variables allows the global optimum to be derived analytically using elementary trigonometric identities. Closed-form solutions have also been derived in a restricted phase-slope regime, i.e., $|a|\le {4\pi}/[{4D_x-(N_r-2)\lambda}]$ \cite{AF_MA}. {For even $N_r$, when
		$D_x\ge (N_r-\tfrac{1}{2})\lambda$, additional closed-form 
		solutions can be obtained using the replicated-subarray construction
		in \cite{ZeroForcing}.}}
These special cases are useful for intuition and validation, but they do not cover the general regime with larger arrays or larger apertures. Therefore, we next develop a general-purpose first-order method and prove that it converges almost surely to a local minimizer rather than merely to a stationary point.

The first step is to remove a redundant degree of freedom. The objective in \eqref{eq:special_equiv_x_problem} satisfies the common-shift invariance property, i.e., 
\begin{equation}
	\label{eq:pa_shift_invariance}
	f_a(\mathbf{x}+c\mathbf{1})
	=
	f_a(\mathbf{x}),
	\qquad
	\forall c\in\mathbb{R}.
\end{equation}
Hence, for any feasible $\mathbf{x}$, the shifted vector
$
	\tilde{\mathbf{x}}
	=
	\mathbf{x}-x_1\mathbf{1}
$
remains feasible, satisfies $\tilde{x}_1=0$, and has the same objective value. This observation suggests using the first antenna as a reference position. Define
$
	\mathbf{y}
	=
	[y_1,\ldots,y_{N_r-1}]^T
	\triangleq
	[x_2,\ldots,x_{N_r}]^T,
$
then the steering-coherence minimization problem \eqref{eq:special_equiv_x_problem} can be reduced as
\begin{equation}
	\label{eq:reduced_problem}
	\min_{\mathbf{y}\in\mathcal{Y}}
	\quad
	F_a(\mathbf{y})
	=
	\left|
	1+\sum_{i=1}^{N_r-1}e^{j a y_i}
	\right|^2,
\end{equation}
where
\begin{equation}
	\label{eq:Y_set}
	\mathcal{Y}
	=
	\left\{
	\mathbf y\in\mathbb R^{N_r-1}
	\middle|
	\begin{array}{l}
		d\le y_1, d\le y_i-y_{i-1},y_{N_r-1}\le D_x,\\
		 \quad\quad \quad i=2,\ldots,N_r-1
	\end{array}
	\right\}.
\end{equation}
To make the role of this reduction explicit, we first state the following lemma.

\begin{Lemma}
	\label{lem:reduced_geometry}
	The reduced problem \eqref{eq:reduced_problem} is equivalent to the original steering-coherence minimization problem \eqref{eq:special_equiv_x_problem} up to the common-shift invariance of $f_a(\mathbf{x})$. In particular, every local minimizer $\mathbf{y}^\star$ of \eqref{eq:reduced_problem} gives a local minimizer $\mathbf{x}^\star=[0,(\mathbf{y}^\star)^T]^T$ of \eqref{eq:special_equiv_x_problem}. Moreover, the feasible set $\mathcal{Y}$ satisfies the linear independence constraint qualification (LICQ) at every feasible point.
\end{Lemma}

\begin{proof}[Proof Sketch]
First, by the common-shift invariance in \eqref{eq:pa_shift_invariance}, we can set $x_1=0$
without loss of optimality. The original and reduced problems are
therefore equivalent up to a common translation, so a local minimizer
of \eqref{eq:reduced_problem} yields a local minimizer of \eqref{eq:special_equiv_x_problem}.
	Second, for LICQ, the constraint normals in $\mathcal{Y}$ can be linearly
	dependent only if all constraints are active. This would force all adjacent antenna spacings to equal $d$ and the last antenna to be
	located at $D_x$, yielding $D_x=(N_r-1)d$, which contradicts
	$D_x>(N_r-1)d$. Therefore, the active constraint normals are
	linearly independent at every feasible point.
	The complete proof is provided in Appendix~A of the supplementary
	material.
\end{proof}

To study the non-minimizing critical points of the constrained problem,
we introduce the following extended objective:
\begin{equation}
	\label{eq:pa_Psi_def}
	\Psi_a(\mathbf{y})
	\triangleq
	F_a(\mathbf{y})+\delta_{\mathcal{Y}}(\mathbf{y}),
\end{equation}
where $\delta_{\mathcal{Y}}$ is the indicator function of
$\mathcal{Y}$, i.e., $\delta_{\mathcal{Y}}(\mathbf{y})=0$ for
$\mathbf{y}\in\mathcal{Y}$ and
$\delta_{\mathcal{Y}}(\mathbf{y})=+\infty$ otherwise.
Let the constraints defining $\mathcal{Y}$ be written as
$c_i(\mathbf{y})\leq 0$, where
\begin{equation}
	\label{eq:pa_constraints_c}
	\left\{
	\begin{aligned}
		c_1(\mathbf y)&=d-y_1,\\
		c_i(\mathbf y)&=d+y_{i-1}-y_i,\quad i=2,\ldots,N_r-1,\\
		c_{N_r}(\mathbf y)&=y_{N_r-1}-D_x .
	\end{aligned}
	\right.
\end{equation}
The active set at $\mathbf{y}$ is
\begin{equation}
	\label{eq:pa_active_set_def}
	A(\mathbf{y})
	=
	\{i:c_i(\mathbf{y})=0\}.
\end{equation}
For a feasible point $\bar{\mathbf{y}}\in\mathcal{Y}$, define
\begin{equation}
	\label{eq:pa_active_face_def}
	\mathcal{M}_{A}
	=
	\left\{
	\mathbf{y}\in\mathbb{R}^{N_r-1}:
	c_i(\mathbf{y})=0,\ \forall i\in A(\bar{\mathbf{y}})
	\right\}
\end{equation}
as the affine set induced by $A(\bar{\mathbf y})$.
By Lemma~\ref{lem:reduced_geometry}, the gradients of these active
constraints, i.e., $\{\nabla c_i:i\in A(\bar{\mathbf{y}})\}$,   are linearly independent, so $\mathcal{M}_A$ is a smooth
affine manifold of dimension
$N_r-1-\lvert A(\bar{\mathbf{y}})\rvert$.
Moreover, since $c_j(\bar{\mathbf{y}})<0$ for every
$j\notin A(\bar{\mathbf{y}})$, the inactive constraints remain
inactive in a sufficiently small neighborhood of $\bar{\mathbf{y}}$.
Hence, near $\bar{\mathbf{y}}$, $\mathcal{M}_A$ describes the face of
$\mathcal{Y}$ determined by the active constraints.
With the local face structure established, we next introduce three definitions used to identify non-minimizing critical points as active strict saddles.

\begin{Definition}[Critical point and strict complementarity \cite{book1}]
	\label{def:critical_sc}
	A point $\bar{\mathbf{y}}\in\mathcal{Y}$ is called a critical point of $\Psi_a$ if
	\begin{equation}
		\label{eq:pa_critical_condition}
		\mathbf{0}\in\partial\Psi_a(\bar{\mathbf{y}}),
	\end{equation}
	or equivalently, there exist multipliers $\lambda_i\ge0$ such that
	\begin{equation}
		\label{eq:pa_kkt_condition}
		\nabla F_a(\bar{\mathbf{y}})
		+
		\sum_{i\in A(\bar{\mathbf{y}})}
		\lambda_i\nabla c_i(\bar{\mathbf{y}})
		=
		\mathbf{0}.
	\end{equation}
	It satisfies strict complementarity if every active constraint has a strictly positive multiplier, i.e.,
	$
		\lambda_i>0,
		\forall i\in A(\bar{\mathbf{y}}).
	$
\end{Definition}

%Strict complementarity means that all active constraints are genuinely needed to balance the gradient at the critical point. In polyhedral problems, this condition is important because it prevents a critical point from lying ambiguously on a larger face with zero multiplier support. In our analysis, this property is used to prove that the correct active face can be identified as an active manifold.

\begin{Definition}[Active manifold \cite{ProximalAvoidsActiveStrictSaddles}]
	\label{def:active_manifold}
	Let $\bar{\mathbf{y}}$ be a critical point of $\Psi_a$. A $C^2$ embedded manifold $\mathcal{M}$ containing $\bar{\mathbf{y}}$ is called an active manifold of $\Psi_a$ at $\bar{\mathbf{y}}$ if the following two conditions hold:
	\begin{enumerate}
		\item Smoothness: the restriction $\Psi_a|_{\mathcal{M}}$ is $C^2$ around $\bar{\mathbf{y}}$.
		\item Sharpness: there exists a neighborhood $\mathcal{U}$ of $\bar{\mathbf{y}}$ such that
		\begin{equation}
			\label{eq:pa_active_manifold_sharpness}
			\inf
			\left\{
			\| \mathbf v\|: \mathbf v \in 
\partial\Psi_a(\mathbf{y}), 
			\mathbf{y}\in \mathcal{U} \setminus \mathcal{M}
			\right\}
			>
			0.
		\end{equation}
	\end{enumerate}
\end{Definition}

%The smoothness condition says that the nonsmooth constrained problem becomes a smooth problem once restricted to the correct manifold. The sharpness condition says that points near $\bar{\mathbf{y}}$ but outside this manifold cannot be approximately stationary. Therefore, approximate stationarity near $\bar{\mathbf{y}}$ forces the iterates to lie on the active manifold. In our polyhedral setting, when strict complementarity holds, the active manifold is exactly the active face
%\begin{equation}
%	\label{eq:pa_active_face_def}
%	\mathcal{M}_A
%	=
%	\{\mathbf{y}\in\mathcal{Y}:c_i(\mathbf{y})=0,\ i\in A\}.
%\end{equation}

\begin{Definition}[Active strict saddle \cite{ProximalAvoidsActiveStrictSaddles}]
	\label{def:active_strict_saddle}
	A critical point $\bar{\mathbf{y}}$ of $\Psi_a$ is called an active strict saddle if there exists an active manifold $\mathcal{M}$ of $\Psi_a$ at $\bar{\mathbf{y}}$ and a nonzero tangent direction $\mathbf{v}\in T_{\mathcal{M}}(\bar{\mathbf{y}})$ such that
	\begin{equation}
		\label{eq:pa_active_strict_saddle_def}
		\mathbf{v}^{T}
		\nabla_{\mathcal{M}}^{2}
		\Psi_a(\bar{\mathbf{y}})
		\mathbf{v}
		<
		0.
	\end{equation}
\end{Definition}

%An active strict saddle is the constrained analogue of a strict saddle in smooth unconstrained optimization. The tangent direction $\mathbf{v}$ is feasible to first order within the active manifold, and the negative curvature in \eqref{eq:pa_active_strict_saddle_def} means that the objective can be decreased to second order along the correct face. Such a point is not only nonoptimal; it is dynamically unstable for the damped projected-gradient map introduced later.

An active strict saddle is locally unstable because the negative-curvature direction along its active manifold enables projected-gradient dynamics to escape from it.
We therefore show that every non-minimizing critical point of $\Psi_a$ is an active strict saddle with the following two steps.
First, the local face determined by the active constraints needs to be identified as an active manifold, which requires smoothness and sharpness of the extended objective $\Psi_a$ on this face. The second is a facewise curvature statement, which requires the restriction of $F_a$ to that face to admit a direction of strictly negative curvature. Lemma~\ref{lem:reduced_geometry} already gives the smoothness part through LICQ. The next lemma shows that, except for an at most countable set of phase slopes, strict complementarity supplies the sharpness part and identifies the active face as the active manifold.

\begin{Lemma}[Generic strict complementarity and active-manifold identification]
	\label{lem:generic_strict_complementarity}
	There exists an at most countable set
	$\mathcal{B}_{\mathrm{sc}}\subset\mathbb{R}$, with
	$0\in\mathcal{B}_{\mathrm{sc}}$, such that, for every
	$a\notin\mathcal{B}_{\mathrm{sc}}$, every critical point
	$\bar{\mathbf{y}}$ of $\Psi_a$ that is not a local minimizer of
	\eqref{eq:reduced_problem} satisfies strict complementarity.
Moreover, for any such critical point, the affine manifold
$\mathcal{M}_{A}$ defined in \eqref{eq:pa_active_face_def} is a
$C^\infty$ active manifold of $\Psi_a$ at $\bar{\mathbf{y}}$.
More precisely, the restriction $\Psi_a|_{\mathcal{M}_{A}}$ is
$C^\infty$ around $\bar{\mathbf{y}}$, and the sharpness condition
\eqref{eq:pa_active_manifold_sharpness} holds.
Consequently, if $a$ is drawn from a non-atomic probability
distribution, then these properties hold almost surely.
	
\end{Lemma}

\begin{proof}[Proof Sketch]
By LICQ, the KKT multipliers at any critical point are unique. If strict complementarity fails at a non-minimizing critical point, the corresponding active-constraint configuration imposes a nontrivial real-analytic condition on the phase slope $a$. Such a condition can hold for at most countably many values of $a$, and only finitely many active-constraint configurations are possible. Hence, all exceptional phase slopes form an at most countable set $\mathcal{B}_{\mathrm{sc}}$, and strict complementarity holds whenever $a\notin\mathcal{B}_{\mathrm{sc}}$. 
LICQ also makes the active face \(\mathcal{M}_A\) a smooth affine manifold. Under strict complementarity, every active constraint has a strictly positive multiplier in the unique KKT condition. Any nearby point outside \(\mathcal{M}_A\) releases at least one of these constraints, so the remaining active constraints cannot satisfy the stationarity condition with an arbitrarily small residual. The subgradient norm is therefore bounded away from zero, which gives sharpness and identifies \(\mathcal{M}_A\) as an active manifold of \(\Psi_a\). The complete proof is provided in Appendix~B of the supplementary material.
\end{proof}

The exceptional set $\mathcal{B}_{\mathrm{sc}}$
contains at most countably many phase-slope values $a$ corresponding to special phase configurations for which strict complementarity may fail, and these exceptional cases occur with probability zero under a continuous angular model.
Thus, for the generic case, i.e., $a\notin\mathcal{B}_{\mathrm{sc}}$, Lemma~\ref{lem:generic_strict_complementarity}
allows us to study a non-minimizing boundary critical point by restricted objective $F_a|_{\mathcal{M}_A}$. We next show that such a point always has a negative-curvature direction on $\mathcal{M}_A$.

\begin{Lemma}
	\label{lem:phasor_face_structure}
	Let
	\begin{equation}
		\label{eq:def_H_beta}
		H(\bm{\beta})
		=
		\left|
		c+\sum_{i=1}^{M}\rho_i e^{j a\beta_i}
		\right|^2,
		\qquad
		a\ne0,
		\qquad
		\rho_i>0,
	\end{equation}
	where $c\in\mathbb{C}$ is fixed and
	$\bm{\beta}\in\mathbb{R}^{M}$. Every critical point of $H(\bm{\beta})$ that is
	not a local minimizer of $H(\bm{\beta})$ admits a direction of strictly negative
	curvature. 	
	
	Furthermore, for any affine manifold $\mathcal{M}_A$ determined by a feasible active set, the restriction $F_a|_{\mathcal{M}_A}$ can be written in the form of \eqref{eq:def_H_beta}. Consequently, if $\widetilde{\mathbf{y}}\in\mathcal{M}_A$ is a critical point of $F_a|_{\mathcal{M}_A}$ that is not a local minimizer of $F_a|_{\mathcal{M}_A}$, then there exists a nonzero tangent direction $\mathbf{v}\in T_{\mathcal{M}_A}(\widetilde{\mathbf{y}})$ such that \begin{equation} \begin{aligned} \mathbf{v}^{T} \nabla_{\mathcal{M}_A}^{2} F_a(\widetilde{\mathbf{y}}) \mathbf{v} < 0 . \end{aligned} \end{equation}

\end{Lemma}

\begin{proof}[Proof Sketch]
	A useful geometric view is to regard each term
	$\rho_i e^{ja\beta_i}$ as a vector of fixed length in the complex
	plane, with its direction controlled by $\beta_i$. The objective $H$
	is the squared magnitude of the overall vector sum
	$c+\sum_{i=1}^{M}\rho_i e^{ja\beta_i}$. At a critical point, all
	movable vectors must lie along the direction of this sum, either
	pointing in the same or opposite direction. 
	If this collinear configuration is not a global minimum, the phasor directions can be perturbed to reduce the magnitude of the sum. Since the first-order variation vanishes at a critical point, this implies the existence of a strictly negative-curvature direction.
	The restriction $F_a|_{\mathcal{M}_A}$ has the same geometric form as
	\eqref{eq:def_H_beta}. Hence, the same argument applies on
	$\mathcal{M}_A$, and every non-minimizing critical point admits a
	strictly negative curvature direction tangent to
	$\mathcal{M}_A$. The complete proof is provided in Appendix~C of the
	supplementary material.
\end{proof}

Lemma~\ref{lem:phasor_face_structure} shows that, on any active affine manifold $\mathcal{M}_A$, every critical point of the restricted objective $F_a|_{\mathcal{M}_A}$ that is not locally minimizing admits a strictly negative curvature direction tangent to $\mathcal{M}_A$. 
Together with Lemma~\ref{lem:generic_strict_complementarity}, this yields the following landscape characterization.

\begin{Theorem}[Active strict-saddle landscape]
	\label{thm:active_strict_saddle}
	For every $a\notin\mathcal{B}_{\mathrm{sc}}$, each critical point
	$\bar{\mathbf{y}}$ of $\Psi_a$ is either a local minimizer of
	\eqref{eq:reduced_problem} or an active strict saddle.
\end{Theorem}

\begin{proof}[Proof Sketch]
	Following Fig.~\ref{fig:active_strict_saddle_roadmap}, consider a
	critical point $\bar{\mathbf y}$ that is not a local minimizer.
	Lemma~2 identifies the corresponding $\mathcal{M}_A$ as an active
	manifold of $\Psi_a$. Meanwhile, the KKT condition together with LICQ
	shows that $\bar{\mathbf y}$ is stationary for the restricted objective
	$F_a|_{\mathcal{M}_A}$. Strict complementarity further implies that
	local minimality on $\mathcal{M}_A$ would also give local minimality
	over $\mathcal{Y}$. Hence, $\bar{\mathbf y}$ is stationary but not
	locally minimizing on $\mathcal{M}_A$.	
	Lemma~3 then shows that $F_a|_{\mathcal{M}_A}$ has the free-phasor
	structure and that such a non-minimizing stationary point admits a
	strictly negative-curvature direction within $\mathcal{M}_A$.
	Combining this negative curvature with the active-manifold property of
	$\mathcal{M}_A$ proves that $\bar{\mathbf y}$ is an active strict saddle
	of $\Psi_a$.	
	The complete proof is provided in Appendix~D of the supplementary
	material.
\end{proof}

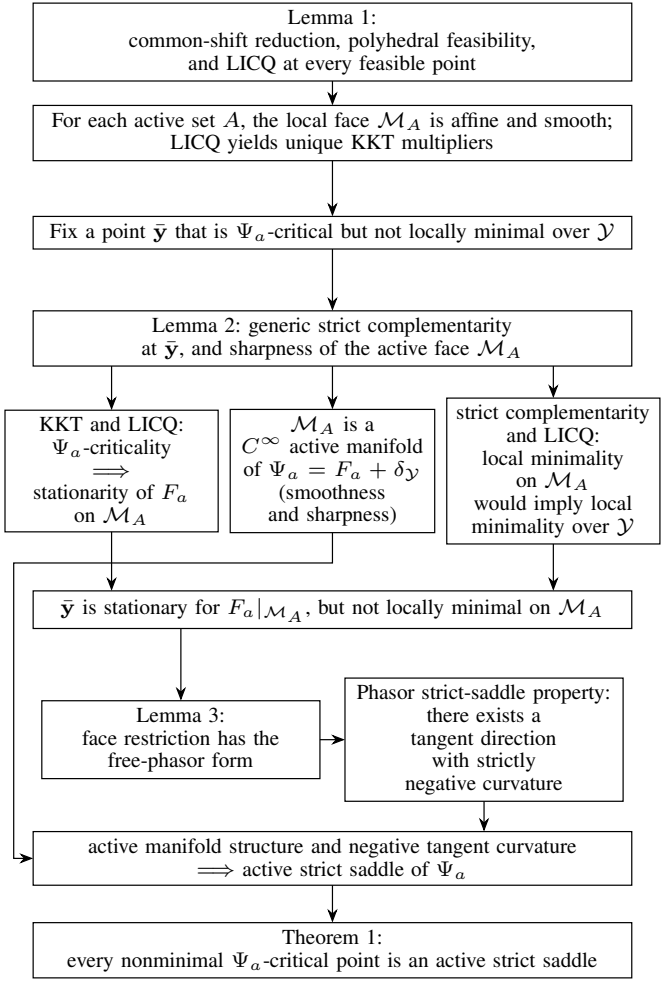
\begin{figure}[t]
	\centering
	\resizebox{0.98\columnwidth}{!}{%
		\begin{tikzpicture}[
			font=\scriptsize,
			box/.style={
				draw,
				align=center,
				inner xsep=3pt,
				inner ysep=2.2pt,
				outer sep=0pt
			},
			wide/.style={box, text width=6.7cm},
			sidebranch/.style={box, text width=2.25cm},
			activebranch/.style={box, text width=2.15cm},
			mid/.style={box, text width=3.0cm},
			arr/.style={-{Stealth[length=1.6mm,width=1.2mm]}, thin}
			]

			\node[wide] (reduction) at (0,0)
			{Lemma~\ref{lem:reduced_geometry}:\\
				common-shift reduction, polyhedral feasibility,\\
				and LICQ at every feasible point};
			
			\node[wide] (faces) at (0,-1.05)
			{For each active set $A$, the local face
				$\mathcal M_A$ is affine and smooth;\\
				LICQ yields unique KKT multipliers};
			
			\node[wide] (hyp) at (0,-2.20)
			{Fix a point $\bar{\mathbf y}$ that is
				$\Psi_a$-critical but not locally minimal over $\mathcal Y$};
			
			\node[wide] (sclemma) at (0,-3.42)
			{Lemma~\ref{lem:generic_strict_complementarity}:
				generic strict complementarity at $\bar{\mathbf y}$,
				and sharpness of the active face $\mathcal M_A$};
			
			\node[sidebranch] (kkt) at (-2.55,-4.95)
			{KKT and LICQ:\\
				$\Psi_a$-criticality\\
				$\Longrightarrow$\\
				stationarity of $F_a$\\
				on $\mathcal M_A$};
			
			\node[activebranch] (active) at (0,-4.95)
			{$\mathcal M_A$ is a\\
				$C^\infty$ active manifold\\
				of $\Psi_a=F_a+\delta_{\mathcal Y}$\\
				(smoothness and sharpness)};
			
			\node[sidebranch] (nonmin) at (2.55,-4.95)
			{strict complementarity and LICQ:\\
				local minimality on $\mathcal M_A$\\
				would imply local minimality over $\mathcal Y$};
			
			\node[wide] (combine) at (0,-6.55)
			{$\bar{\mathbf y}$ is stationary for $F_a|_{\mathcal M_A}$,
				but not locally minimal on $\mathcal M_A$};
			
			\node[mid] (phasor) at (-1.75,-8.05)
			{Lemma~\ref{lem:phasor_face_structure}:\\
				face restriction has the\\
				free-phasor form};
			
			\node[mid] (curv) at (1.75,-8.05)
			{Phasor strict-saddle property:\\
				there exists a tangent direction\\
				with strictly negative curvature};
			
			\node[wide] (activecurv) at (0,-9.42)
			{active manifold structure and negative tangent curvature\\
				$\Longrightarrow$ active strict saddle of $\Psi_a$};
			
			\node[wide] (theorem) at (0,-10.48)
			{Theorem~\ref{thm:active_strict_saddle}:\\
				every nonminimal $\Psi_a$-critical point is an active strict saddle};
			
			\draw[arr] (reduction) -- (faces);
			\draw[arr] (faces) -- (hyp);
			\draw[arr] (hyp) -- (sclemma);
			
			\draw[arr] ([xshift=-2.55cm]sclemma.south) -- (kkt.north);
			\draw[arr] (sclemma.south) -- (active.north);
			\draw[arr] ([xshift=2.55cm]sclemma.south) -- (nonmin.north);
			
			\draw[arr] (kkt.south) -- ([xshift=-2.55cm]combine.north);
			\draw[arr] (nonmin.south) -- ([xshift=2.55cm]combine.north);
			
			\draw[arr] ([xshift=-1.75cm]combine.south) -- (phasor.north);
			\draw[arr] (phasor.east) -- (curv.west);
			
			\draw[arr] (curv.south) -- ([xshift=1.75cm]activecurv.north);
			\draw[arr] (active.south) -- ++(0,-0.35) -- ++(-3.68,0) |- (activecurv.west);
			
			\draw[arr] (activecurv) -- (theorem);
			
		\end{tikzpicture}%
	}
	\caption{Proof structure of Theorem~\ref{thm:active_strict_saddle}.}
	\label{fig:active_strict_saddle_roadmap}

\end{figure}

Theorem~\ref{thm:active_strict_saddle} shows that, except for an exceptional set of phase slopes, i.e.,
$a\notin\mathcal{B}_{\mathrm{sc}}$,
the reduced problem \eqref{eq:reduced_problem} has no other type of non-minimizing critical point, which means that every such point is an active strict saddle. Hence, the only stable candidates for DPGD convergence are local minima, provided that the algorithm avoids active strict saddles. 
%The key point is that Lemma~\ref{lem:generic_strict_complementarity} identifies the relevant active face $\mathcal{M}_A$ as the active manifold of $\Psi_a$, while Lemma~\ref{lem:phasor_face_structure} supplies the negative curvature on this manifold. The two bridge steps in the proof ensure that the curvature lemma applies to the same point, i.e., KKT converts $\Psi_a$-criticality into stationarity of $F_a|_{\mathcal{M}_A}$, and LICQ together with strict complementarity transfers nonminimality from $\mathcal{Y}$ to $\mathcal{M}_A$. Since $\mathcal{B}_{\mathrm{sc}}$ is at most countable, the dichotomy holds almost surely whenever the phase slope $a$ has a non-atomic distribution. In particular, this applies to any angular model for which $a=k(\sin\theta_1-\sin\phi_1)$ is a non-atomic random variable.

\subsection{Proposed DPGD algorithm}
\label{subsec:dpgd_algorithm}
The proposed DPGD follows a damped projected-gradient structure. Starting from a feasible point, each iteration first takes a gradient step on $F_a$, projects the resulting point onto $\mathcal{Y}$ to satisfy the spacing and aperture constraints, and then applies damping between the current and projected points. As shown below, the projection can be implemented efficiently by PAVA. We further show that the fixed points of this iteration coincide with the critical points of $\Psi_a$, which connects the algorithm with the landscape result established above. Its almost-sure convergence to a local minimizer is analyzed in the next subsection.

First, we define the phasor sum $
	S_a(\mathbf{y})
	=
	1+\sum_{i=1}^{N_r-1}e^{j a y_i}
$,
then $F_a(\mathbf{y})=\lvert S_a(\mathbf{y})\rvert^2$, and
\begin{equation}
	\label{eq:grad_Fa}
	\left[\nabla F_a(\mathbf{y})\right]_i
	=
	-2a\,\operatorname{Im}
	\left(
	e^{ja y_i}S_a^*(\mathbf{y})
	\right),
	i=1,\ldots,N_r-1 .
\end{equation}
A valid global Lipschitz constant of $\nabla F_a$ is
\begin{equation}
	\label{eq:pa_La_def}
	L_a
	=
	4N_r a^2,
\end{equation}
i.e., $
	\left\|
	\nabla F_a(\mathbf{u})-\nabla F_a(\mathbf{v})
	\right\|
	\le
	L_a
	\left\|
	\mathbf{u}-\mathbf{v}
	\right\|,
	\forall \mathbf{u},\mathbf{v}\in\mathbb{R}^{N_r-1}.
$
Thus, $L_a$ quantifies the global curvature of the smooth part $F_a$ and
determines the admissible range of the projected-gradient stepsize.

Given a stepsize $\mu>0$ and a damping parameter
$\eta\in(0,1)$, define the projected-gradient map and its damped
counterpart as
\begin{equation}
	\label{eq:dpgd_maps}
	\begin{aligned}
		G_{\mu,a}(\mathbf y)
		&\triangleq
		\Pi_{\mathcal Y}
		\left(
		\mathbf y-\mu\nabla F_a(\mathbf y)
		\right),\\
		T_{\mu,\eta,a}(\mathbf y)
		&\triangleq
		(1-\eta)\mathbf y
		+
		\eta G_{\mu,a}(\mathbf y).
	\end{aligned}
\end{equation}
The DPGD iteration can thus be written as
\begin{equation}
	\label{eq:dpgd_update_compact}
	\mathbf y^{k+1}
	=
	T_{\mu,\eta,a}(\mathbf y^k).
\end{equation}
Since $\mathcal Y$ is convex,
$G_{\mu,a}(\mathbf y^k)\in\mathcal Y$ and
$T_{\mu,\eta,a}(\mathbf y^k)\in\mathcal Y$ whenever
$\mathbf y^k\in\mathcal Y$. Moreover, since $\eta>0$,
\begin{equation}
	\label{eq:fixed_point_critical_equiv}
	\begin{aligned}
		\mathbf y=T_{\mu,\eta,a}(\mathbf y)
		&\Longleftrightarrow
		\mathbf y=G_{\mu,a}(\mathbf y)\\
		&\Longleftrightarrow
		\mathbf 0
		\in
		\nabla F_a(\mathbf y)
		+
		N_{\mathcal Y}(\mathbf y)=\partial\Psi_a(\mathbf y).
	\end{aligned}
\end{equation}
Hence, $G_{\mu,a}$ and $T_{\mu,\eta,a}$ have the same fixed
points, which are exactly the critical points of $\Psi_a$.
For any
$\mathbf{q}\in\mathbb{R}^{N_r-1}$, set
$b_i=q_i-id$ and write $r_i=y_i-id$. Then the constraints in
\eqref{eq:Y_set} are transformed into the following bounded isotonic constraints:
\begin{equation}
	\label{eq:isotonic_constraints}
	0
	\le
	r_1
	\le
	r_2
	\le
	\cdots
	\le
	r_{N_r-1}
	\le
	D_x-(N_r-1)d .
\end{equation}
Therefore, $\Pi_{\mathcal{Y}}(\mathbf{q})$ can be obtained by projecting
$\mathbf{b}=[b_1,\ldots,b_{N_r-1}]^T$ onto the bounded isotonic set in
\eqref{eq:isotonic_constraints}, followed by the inverse change of
variables $y_i=r_i+id$. This bounded isotonic regression problem can be
solved by the PAVA with
$O(N_r)$ complexity~\cite{PTRSO,PAVA}.

\begin{algorithm}[t]
	\caption{Proposed DPGD Algorithm}
	\label{alg:DPGD}
	\renewcommand{\algorithmicrequire}{\textbf{Input:}}
	\renewcommand{\algorithmicensure}{\textbf{Output:}}
	\begin{algorithmic}[1]
		\footnotesize
		\REQUIRE desired/jammer AoA pair $(\theta_1,\phi_1)$, wavenumber $k$,
		$N_r$, $d$, $D_x$, stepsize $\mu$, damping parameter $\eta$,
		tolerance $\epsilon$, maximum iteration number $K_{\max}$, and
		feasible initial point $\mathbf{y}^{0}\in\mathcal{Y}$.
		\ENSURE APV $\mathbf{x}_{\star}$ and MVDR beamformer
		$\mathbf{w}_{\mathrm{opt}}(\mathbf{x}_{\star})$.
		\STATE Set $a=k(\sin\theta_1-\sin\phi_1)$ and $m=0$.
%		\IF{$a=0$}
%		\STATE Set $\mathbf{x}_{\star}=[0,(\mathbf{y}^{0})^T]^T$.
%		\STATE Compute $\mathbf{w}_{\mathrm{opt}}(\mathbf{x}_{\star})$
%		by \eqref{eq:wopt_nlos}.
%		\RETURN $\mathbf{x}_{\star}$ and
%		$\mathbf{w}_{\mathrm{opt}}(\mathbf{x}_{\star})$.
%		\ENDIF
		\WHILE{$m<K_{\max}$}
		\STATE Compute $\nabla F_a(\mathbf{y}^{m})$ by
		\eqref{eq:grad_Fa}.
		\STATE Compute $\mathbf{y}^{m+1}$
		according to \eqref{eq:dpgd_update_compact}, using bounded
		PAVA for $\Pi_{\mathcal{Y}}$.
		\IF{$\|\mathbf{y}^{m+1}-\mathbf{y}^{m}\|_2\le\epsilon$}
		\STATE $m\leftarrow m+1$.
		\STATE \textbf{break}
		\ENDIF
		\STATE $m\leftarrow m+1$.
		\ENDWHILE
		\RETURN $\mathbf{x}_{\star}=[0,(\mathbf{y}^{m})^T]^T$ and compute $\mathbf{w}_{\mathrm{opt}}(\mathbf{x}_{\star})$
		by \eqref{eq:wopt_nlos}.
	\end{algorithmic}
\end{algorithm}

Algorithm~\ref{alg:DPGD} summarizes the implementation. Each
iteration requires one gradient evaluation and one projection onto
$\mathcal Y$. The phasor sum $S_a(\mathbf y)$ and the gradient
$\nabla F_a(\mathbf y)$ can be evaluated in
$\mathcal O(N_r)$ operations, while the bounded PAVA projection also
has $\mathcal O(N_r)$ complexity. Hence, the per-iteration
complexity of Algorithm~\ref{alg:DPGD} is $\mathcal O(N_r)$, and the APV optimization over $K$
iterations requires $\mathcal O(KN_r)$ operations. After obtaining
the APV, the MVDR beamformer is recovered from the rank-one
Sherman-Morrison expression in
\eqref{eq:special_Rinv_simplified} with
$\mathcal O(N_r)$ complexity. Therefore, the overall computational
complexity of Algorithm~\ref{alg:DPGD} is
$
	\mathcal O(KN_r).
$

\subsection{Almost-Sure Convergence to Local Minima}
\label{subsec:dpgd_convergence}

We next show that the DPGD algorithm almost surely avoids the active strict saddles characterized above. The fixed-point relation \eqref{eq:fixed_point_critical_equiv} shows that any limit point of the iteration, if it exists, must be a critical point of $\Psi_a$. The role of damping is to make the iteration map invertible without changing its fixed points, which is needed for the saddle-avoidance result. We use the following two standard notions from dynamical systems \cite{ProximalAvoidsActiveStrictSaddles}.

\begin{Definition}[Lipeomorphism \cite{ProximalAvoidsActiveStrictSaddles}]
	\label{def:lipeomorphism}
	A map $T:\mathcal{U}\rightarrow\mathcal{V}$ between subsets of
	Euclidean spaces is called a lipeomorphism if it is bijective and both
	$T$ and $T^{-1}$ are Lipschitz continuous.
\end{Definition}

\begin{Definition}[Local center-stable manifold \cite{ProximalAvoidsActiveStrictSaddles}]
	\label{def:center_stable_manifold}
	Let $G$ be a $C^1$ local diffeomorphism around a fixed point
	$\bar{\mathbf{y}}$. A local center-stable manifold of $G$ at
	$\bar{\mathbf{y}}$ is a local embedded manifold
	$\mathcal{W}^{cs}_{\mathrm{loc}}$ passing through
	$\bar{\mathbf{y}}$ such that
	\begin{equation}
		\label{eq:pa_center_stable_tangent}
		T_{\bar{\mathbf{y}}}\mathcal{W}^{cs}_{\mathrm{loc}}
		=
		E_{\le1}
		\left(
		\nabla G(\bar{\mathbf{y}})
		\right),
	\end{equation}
	where $E_{\le1}(\nabla G(\bar{\mathbf{y}}))$ denotes the direct sum
	of the generalized eigenspaces associated with eigenvalues of modulus
	at most one.
	Moreover, there exists a sufficiently small neighborhood
	$\mathcal{U}$ of $\bar{\mathbf{y}}$ such that, for any
	$\mathbf{y}\in\mathcal{U}$ satisfying
	$
	G^k(\mathbf{y})\in\mathcal{U},
	 \forall k\ge0,
	$
	we have
	$\mathbf{y}\in\mathcal{W}^{cs}_{\mathrm{loc}}$.
\end{Definition}

The notions above allow us to analyze whether DPGD can converge to the active strict saddles characterized above. The following theorem establishes the resulting local-minimum convergence guarantee under suitable stepsize and damping conditions.
%The proof of the next theorem has two parts. First, Appendix~E establishes that the entire DPGD sequence converges to a critical point of $\Psi_a$. Second, if this critical point is not a local minimizer, Theorem~\ref{thm:active_strict_saddle} identifies it as an active strict saddle. On the corresponding active manifold, the negative tangent curvature makes the fixed point of $T_{\mu,\eta,a}$ unstable, and the local center-stable manifold theorem confines nearby initializations converging to it to a set of Lebesgue measure zero. Under the stepsize and damping conditions below, $T_{\mu,\eta,a}$ is a lipeomorphism, so the backward images of this local exceptional set also have measure zero. Consequently, an absolutely continuous initialization almost surely avoids all non-minimizing critical points.

\begin{Theorem}[Almost-sure convergence of DPGD to a local minimizer]
	\label{thm:dpgd_convergence}
	Suppose that the random variable
	$a=k(\sin\theta_1-\sin\phi_1)$ is absolutely continuous with respect
	to the Lebesgue measure.\footnote{This assumption holds for common continuous angular models, e.g., when $\theta_1$ and $\phi_1$ are independently drawn from distributions
		with continuous densities.} 
	For each realization
	$a\notin\mathcal{B}_{\mathrm{sc}}$, choose
	\begin{equation}
		\label{eq:stepsize_condition}
		0<\mu<\frac{2}{5L_a},
		\qquad
		0<\eta<\frac{1}{1+2\mu L_a},
	\end{equation}
	where $L_a$ is defined in \eqref{eq:pa_La_def}. If
	$\mathbf{y}^{0}$ is drawn independently of $a$ from a distribution on
	$\mathcal{Y}$ that is absolutely continuous with respect to the
	Lebesgue measure on $\mathbb{R}^{N_r-1}$, then the DPGD sequence
	generated by \eqref{eq:dpgd_update_compact} converges almost surely
	to a local minimizer $\mathbf{y}_{\star}$ of
	\eqref{eq:reduced_problem}. Consequently,
	$\mathbf{x}_{\star}=[0,\mathbf{y}_{\star}^{T}]^{T}$ is a local
	minimizer of the original steering-coherence minimization problem
	\eqref{eq:special_equiv_x_problem}.
\end{Theorem}

\begin{proof}[Proof Sketch]
	Since $\mathcal{B}_{\mathrm{sc}}$ is at most countable and $a$
	is absolutely continuous, $a\notin\mathcal{B}_{\mathrm{sc}}$
	almost surely. By Theorem~1, every
	non-minimizing critical point is an active strict saddle. For each
	such saddle, the corresponding local center-stable manifold
	$\mathcal{W}_{\mathrm{loc}}^{\mathrm{cs}}$ of the DPGD map is
	lower-dimensional, and any trajectory converging to the saddle
	must eventually lie on this manifold. The damping condition in
	\eqref{eq:stepsize_condition} makes $T_{\mu,\eta,a}$ a lipeomorphism, so the inverse
	images of this local exceptional set under earlier iterations also
	have measure zero. Hence, an absolutely continuous random
	initialization converges to an active strict saddle with probability
	zero \cite{ProximalAvoidsActiveStrictSaddles}.
	The stepsize condition in \eqref{eq:stepsize_condition}, together with the compactness of
	$\mathcal{Y}$ and the Kurdyka-{\L}ojasiewicz property \cite{ProximalAvoidsActiveStrictSaddles}, ensures
	that the entire DPGD sequence converges to a critical point of
	$\Psi_a$.
	Thus, the limiting critical point is almost surely a local
	minimizer. Finally, Lemma~1 transfers the result from
	$\mathbf y^\star$ to
	$\mathbf x^\star=[0,(\mathbf y^\star)^T]^T$.
	The complete proof is provided in Appendix~E of the supplementary
	material.
\end{proof}

\section{General Multipath Multi-Jammer Case}
\label{sec:exact_reformulation_general}

In this section, we consider the general problem in \eqref{eq:P0_nlos}, which allows multiple propagation paths on the legitimate link and/or multiple jammers. The objective is highly nonconvex because the antenna positions enter the desired multipath channel and all jammer channels through complex exponentials. 
The main difficulty is that the APV appears on both the desired channel
vector and the interference-plus-noise covariance in the MVDR
objective, i.e., each coordinate $x_n$ changes the $n$-th entry of $\mathbf h_0(\mathbf x)$ and the $n$-th entries of all $\mathbf g_i(\mathbf x)$ simultaneously, and hence also changes the interference-plus-noise covariance $\mathbf R_{i+n}(\mathbf x)$. 
A direct differentiation of $\mathbf h_0^H(\mathbf x)\mathbf R_{i+n}^{-1}(\mathbf x)\mathbf h_0(\mathbf x)$ would require
\begin{equation}
	\label{eq:inverse_derivative_general}
	\frac{\partial \mathbf R_{i+n}^{-1}}{\partial x_n}
	=
	-\mathbf R_{i+n}^{-1}
	\frac{\partial \mathbf R_{i+n}}{\partial x_n}
	\mathbf R_{i+n}^{-1}
\end{equation}
for every antenna coordinate, which involves repeated $N_r \times N_r$ covariance matrix inversion. To avoid this, we reformulate the objective in \eqref{eq:P0_nlos} as an equivalent  ridge LS problem. Since the ridge subproblem has a unique minimizer, Danskin's theorem \cite{danskin1966theory} yields the objective gradient without differentiating the covariance inverse, while the required matrix solve is reduced to an $I\times I$ positive-definite system.
Based on this reformulation and the resulting gradient expression, we develop the proposed SAPG algorithm, which uses a bounded slack-dependent metric to bias the projected step toward the tangent space of near-active constraints, thereby incorporating the active manifold intuition behind saddle avoidance.
Finally, we show that the proposed APV optimization framework, developed for MVDR reception (or equivalently, LMMSE reception), extends naturally to ZF reception.

\subsection{Ridge LS Reformulation}
\label{subsec:exact_gradient_general}
\label{subsec:low_rank_reduction_general}
\label{subsec:ridge_residual_reformulation_general}
\label{subsec:danskin_gradient_general}

This subsection reformulates the APV-only objective in
\eqref{eq:P0_nlos} as an equivalent  ridge LS problem and then derives
its gradient via Danskin's theorem \cite{hoerl1970ridge, danskin1966theory}. 
The key observation is that $\mathbf R_{i+n}(\mathbf x)$ is a diagonal noise matrix plus an $I$-rank jammer covariance. Collect the scaled jammer channels in
\begin{equation}
	\label{eq:V_def_nlos_rev}
	\mathbf V(\mathbf x)
	\triangleq
	\left[
	\sigma_1\mathbf g_1(\mathbf x),
	\ldots,
	\sigma_I\mathbf g_I(\mathbf x)
	\right]
	\in\mathbb C^{N_r\times I},
\end{equation}
then we have
\begin{equation}
	\label{eq:R_compact_nlos_rev}
	\mathbf R_{i+n}(\mathbf x)
	=
	\sigma_n^2\mathbf I_{N_r}
	+
	\mathbf V(\mathbf x)\mathbf V^H(\mathbf x).
\end{equation}
Instead of forming the $N_r\times N_r$ inverse directly, first define
\begin{equation}
	\label{eq:C_b_def_nlos_rev}
	\begin{aligned}
		\mathbf C(\mathbf x)
		&\triangleq
		\sigma_n^2\mathbf I_I+
		\mathbf V^H(\mathbf x)\mathbf V(\mathbf x),\\
		\mathbf b(\mathbf x)
		&\triangleq
		\mathbf V^H(\mathbf x)\mathbf h_0(\mathbf x).
	\end{aligned}
\end{equation}
Since $\sigma_n^2>0$, $\mathbf C(\mathbf x)$ is Hermitian positive definite for every $\mathbf x$. The Woodbury identity gives
\begin{equation}
	\label{eq:woodbury_nlos_rev}
	\mathbf R_{i+n}^{-1}(\mathbf x)
	=
	\frac{1}{\sigma_n^2}
	\left(
	\mathbf I_{N_r}
	-
	\mathbf V(\mathbf x)
	\mathbf C^{-1}(\mathbf x)
	\mathbf V^H(\mathbf x)
	\right).
\end{equation}
Substituting \eqref{eq:woodbury_nlos_rev} into $g(\mathbf x)$ gives
\begin{equation}
	\label{eq:g_reduced_nlos_rev}
	g(\mathbf x)
	=
	\frac{1}{\sigma_n^2}
	\left(
	\|\mathbf h_0(\mathbf x)\|_2^2
	-
	\mathbf b^H(\mathbf x)
	\mathbf C^{-1}(\mathbf x)
	\mathbf b(\mathbf x)
	\right).
\end{equation}
Thus, the objective in \eqref{eq:P0_nlos} can be evaluated through the $I\times I$ positive-definite system associated with $\mathbf C(\mathbf x)$, rather than through an $N_r\times N_r$ covariance inverse.

The reduced expression in \eqref{eq:g_reduced_nlos_rev} is already useful for objective evaluation, but directly differentiating
$\mathbf b^H(\mathbf x)\mathbf C^{-1}(\mathbf x)\mathbf b(\mathbf x)$
would still introduce derivatives of $\mathbf C^{-1}(\mathbf x)$. 
To avoid this complication and obtain a cleaner gradient formula, we further rewrite \eqref{eq:g_reduced_nlos_rev} as a normalized value function of a ridge LS problem as follows:
\begin{equation}
	\label{eq:J_def_nlos_rev}
	\begin{aligned}
		J(\mathbf x,\boldsymbol\nu)
		\triangleq
		\|\mathbf V(\mathbf x)\boldsymbol\nu
		-
		\mathbf h_0(\mathbf x)\|_2^2
		+
		\sigma_n^2\|\boldsymbol\nu\|_2^2,
		\,
		\boldsymbol\nu\in\mathbb C^I.
	\end{aligned}
\end{equation}
Since $\sigma_n^2>0$, $J(\mathbf x,\boldsymbol\nu)$ is strongly convex in $\boldsymbol\nu$, and its unique minimizer satisfies
\begin{equation}
	\label{eq:nu_star_eq_nlos_rev}
	\boldsymbol\nu^\star(\mathbf x)
	=
	\mathbf C^{-1}(\mathbf x)\mathbf b(\mathbf x).
\end{equation}
Substituting $\boldsymbol\nu^\star(\mathbf x)$ back into \eqref{eq:J_def_nlos_rev} yields
\begin{equation}
	\label{eq:g_equals_Jmin_nlos_rev}
	F(\mathbf x)
	\triangleq\min_{\boldsymbol\nu\in\mathbb C^I}
	J(\mathbf x,\boldsymbol\nu)
	=
	\|\mathbf h_0(\mathbf x)\|_2^2
	-
	\mathbf b^H(\mathbf x)
	\mathbf C^{-1}(\mathbf x)
	\mathbf b(\mathbf x).
\end{equation}
Therefore, we have $g(\mathbf x)=F(\mathbf x)/\sigma_n^2$, and maximizing $g(\mathbf x)$ over $\mathcal X$ is equivalent to maximizing $F(\mathbf x)$ over $\mathcal X$. 
The value-function form in \eqref{eq:g_equals_Jmin_nlos_rev} also simplifies the derivative of $F(\mathbf x)$. Since the inner ridge problem has a unique minimizer, the derivative of the optimal value can be obtained by differentiating $J(\mathbf x,\boldsymbol\nu)$ with respect to $\mathbf x$ while keeping $\boldsymbol\nu$ fixed at $\boldsymbol\nu^\star(\mathbf x)$. This is precisely the content of Danskin's theorem as follows \cite{danskin1966theory}.

\begin{Theorem}[Gradient of the ridge value function]
	\label{thm:danskin_nlos_rev}
	The value function $F$ is continuously differentiable and,
for every $\mathbf x\in\mathcal X$,
\begin{equation}
	\label{eq:danskin_nlos_rev}
	\nabla_{\mathbf x}F(\mathbf x)
	=
	\nabla_{\mathbf x}J(\mathbf x,\boldsymbol\nu)
	\big|_{\boldsymbol\nu=\boldsymbol\nu^\star(\mathbf x)},
\end{equation}
where $\boldsymbol\nu$ is held fixed when evaluating
$\nabla_{\mathbf x}J$.
\end{Theorem}

\begin{proof}[Proof Sketch]
	Since $\sigma_n^2>0$, $J(\mathbf x,\boldsymbol{\nu})$ is strongly
	convex and coercive in $\boldsymbol{\nu}$ and therefore has the unique
	minimizer $\boldsymbol{\nu}^{\star}(\mathbf x)$ in \eqref{eq:nu_star_eq_nlos_rev}.
	Moreover, $\mathbf V(\mathbf x)$ and $\mathbf h_0(\mathbf x)$ are
	smooth in $\mathbf x$, so the regularity conditions of Danskin's
	theorem are satisfied \cite{RockafellarWets1998,AlcantaraTakedaDanskin}. Applying Danskin's theorem then gives
	\eqref{eq:danskin_nlos_rev}. The complete proof is provided in
	Appendix~F of the supplementary material.
\end{proof}

Consequently, the derivative can be evaluated without differentiating either $\mathbf C^{-1}(\mathbf x)$ or $\boldsymbol\nu^\star(\mathbf x)$.
With $\boldsymbol\nu^\star(\mathbf x)$ obtained from \eqref{eq:nu_star_eq_nlos_rev}, define the residual
\begin{equation}
	\label{eq:residual_def_nlos_rev}
	\mathbf r(\mathbf x)
	\triangleq
	\mathbf V(\mathbf x)\boldsymbol\nu^\star(\mathbf x)
	-
	\mathbf h_0(\mathbf x).
\end{equation}
Since $x_n$ affects only the $n$-th row of $\mathbf V(\mathbf x)$ and the $n$-th entry of $\mathbf h_0(\mathbf x)$, \eqref{eq:danskin_nlos_rev} gives
\begin{equation}
	\label{eq:dFdx_basic_nlos_rev}
	\begin{aligned}
		&\frac{\partial F(\mathbf x)}{\partial x_n}\!\!
		= \!-2k\Im\Bigg\{\!\!\!
		\left(
		\sum_{i=1}^{I}\sigma_i\zeta_i e^{jkx_n\sin\phi_i}\nu_i^\star
		-\sum_{\ell=1}^{L}\alpha_\ell e^{jkx_n\sin\theta_\ell} \!\!
		\right)^*
		\\
		&\times\!\!\left(
		\sum_{i=1}^{I}\sigma_i\zeta_i\sin\phi_i e^{jkx_n\sin\phi_i}\nu_i^\star
		-\sum_{\ell=1}^{L}\alpha_\ell\sin\theta_\ell e^{jkx_n\sin\theta_\ell}
		\right)\!\!
		\Bigg\},
	\end{aligned}
\end{equation}
which provides a direct component-wise expression for $\nabla F(\mathbf x)$, and the gradient of the original objective follows as $\nabla g(\mathbf x)=\nabla F(\mathbf x)/\sigma_n^2$.

\subsection{Proposed SAPG Algorithm}
\label{subsec:sapg_general}
By \eqref{eq:g_equals_Jmin_nlos_rev}, maximizing $g(\mathbf x)$ over
$\mathcal X$ is equivalent to
\begin{equation}
	\label{eq:sapg_problem}
	\max_{\mathbf x\in\mathcal X} F(\mathbf x).
\end{equation}
For the general multipath multi-jammer case, the almost-sure
saddle-avoidance result established in the previous section does not
extend directly to \eqref{eq:sapg_problem}. In particular, when
$F(\mathbf x)$ is restricted to an active face, it no longer reduces
to the phasor-sum form
$\left|c+\sum_m\rho_m e^{ja\beta_m}\right|^2$
used in the previous landscape analysis. 
%The ridge solution $\boldsymbol\nu^\star(\mathbf x)$ depends on both the desired
%multipath channel and the jammer channels, which makes the corresponding
%curvature analysis considerably more involved.

Nevertheless, the preceding analysis shows that the geometry of the
active constraints plays an important role in the local behavior of
projected-gradient methods. Near a strictly complementary stationary
point, active and inactive constraints can be distinguished by their
slacks, with smaller slacks indicating constraints that are closer to
being active. 
Motivated by this observation, the proposed SAPG algorithm uses the current slacks to construct a bounded metric for the search direction, instead of identifying the active set explicitly.

For the APV constraints, define the nonnegative corresponding slack functions as
\begin{equation}
	\label{eq:sapg_slacks}
	\left\{
	\begin{aligned}
		s_i(\mathbf{x})&=x_{i+1}-x_i-d,
		\quad i=1,\ldots,N_r-1,\\
		s_{N_r}(\mathbf{x})&=x_1,\\
		s_{N_r+1}(\mathbf{x})&=D_x-x_{N_r},
	\end{aligned}
	\right.
\end{equation}
At iteration $k$, set
$
	s_{\max,k}
	=
	\max_{1\le i\le N_r+1}s_i(\mathbf x_k)$ and $
	\tau_k=\kappa s_{\max,k},
	$
where $0<\kappa\le1$ is a scaling parameter.
SAPG then uses the scale $\tau_k$ to assign each APV constraint the smooth active-set weight
\begin{equation}
	\label{eq:sapg_weights}
	\omega_i(\mathbf{x}_k)
	=
	\frac{1}
	{1+\left(s_i(\mathbf{x}_k)/\tau_k\right)^2},
	\quad
	i=1,\ldots,N_r+1,
\end{equation}
where $0<\omega_i(\mathbf{x}_k)\le1$, so that constraints with smaller slacks are weighted more heavily.
Using the weights in \eqref{eq:sapg_weights}, SAPG defines the following trial direction:
\begin{equation}
	\label{eq:sapg_metric_direction}
	\begin{aligned}
		\mathbf p_k
		=
		\mathbf H_k^{-1}\nabla F(\mathbf{x}_k),
	\end{aligned}
\end{equation}
where $\mathbf H_k=
\mathbf I
+
\gamma_0
\sum_{i=1}^{N_r+1}
\omega_i(\mathbf{x}_k)\mathbf u_i\mathbf u_i^T,
\gamma_0>0$ and $\mathbf u_i^T$ denotes the $i$-th row of $\mathbf U$ in \eqref{eq:U_nlos}. 
Furthermore, $\mathbf H_k$ admits a special tridiagonal structure which costs only linear operations to solve $\mathbf H_k\mathbf p_k=\nabla F(\mathbf x_k)$ \cite{NumericalAlgorithms}.
Since $\mathbf H_k\succ0$, the direction $\mathbf p_k$ is well defined and can be equivalently written as
\begin{equation}
	\label{eq:sapg_variational_direction}
	\begin{aligned}
		\mathbf p_k
		=
		\underset{\mathbf v\in\mathbb R^{N_r}}{\arg\max}
		\Bigg\{
		&\nabla F(\mathbf{x}_k)^T\mathbf v
		-
		\frac{1}{2}\|\mathbf v\|_2^2\\
		&-
		\frac{\gamma_0}{2}
		\sum_{i=1}^{N_r+1}
		\omega_i(\mathbf{x}_k)(\mathbf u_i^T\mathbf v)^2
		\Bigg\},
	\end{aligned}
\end{equation}
where the maximizer of the first two terms, i.e.,  
$\nabla F(\mathbf{x}_k)^T\mathbf v
-
\frac{1}{2}\|\mathbf v\|_2^2$, is the Euclidean gradient $\nabla F(\mathbf{x}_k)$, whereas the additional quadratic term $-
\frac{\gamma_0}{2}
\sum_{i=1}^{N_r+1}
\omega_i(\mathbf{x}_k)(\mathbf u_i^T\mathbf v)^2$ penalizes directions that would rapidly change the slacks of highly weighted constraints. 
Since larger weights correspond to smaller slacks, the resulting direction $\mathbf p_k$ is biased toward the tangent space of the candidate active face indicated by the near-active constraints, which is consistent with the active-manifold motivation above.
To be more specific, 
for any threshold $\omega_0\in(0,1]$, define
$
\mathcal A_k(\omega_0)
=
\{i:\omega_i(\mathbf{x}_k)\ge\omega_0\}$
as the $\omega_0$-near-active set, and let $\mathbf U_{\mathcal A_k(\omega_0)}$ collect the corresponding rows of $\mathbf U$. 
For any nonempty $\mathcal A_k(\omega_0)$, we can obtain $\mathbf H_k\succeq \mathbf I +
\gamma_0\omega_0
\mathbf U_{\mathcal A_k(\omega_0)}^T
\mathbf U_{\mathcal A_k(\omega_0)}$.
Hence, applying Cauchy-Schwarz inequality on \eqref{eq:sapg_metric_direction}, we have $\mathbf p_k^T\mathbf H_k\mathbf p_k
\le \|\mathbf p_k\|_2
\|\nabla F(\mathbf{x}_k)\|_2$, which implies
\begin{equation}
	\label{eq:sapg_tangent_bound}
	\left\|
	\mathbf U_{\mathcal A_k(\omega_0)}
	\mathbf p_k
	\right\|_2
	\le
	\frac{\|\nabla F(\mathbf{x}_k)\|_2}
	{2\sqrt{\gamma_0\omega_0}}.
\end{equation}
When the spacing between two neighboring antennas approaches the minimum value $d$, the corresponding weight increases, so SAPG penalizes their relative displacement and avoids large variations in the already small spacing.
More generally, \eqref{eq:sapg_tangent_bound} shows that larger weights increasingly restrict search directions that would move the current APV away from a near-active constraint boundary.
In particular, when $\omega_0=1$, $\mathcal A_k(1)$ is the exact active set $A$, and the bound becomes
$
\|\mathbf U_{A}\mathbf p_k\|_2
\le
\|\nabla F(\mathbf{x}_k)\|_2/2\sqrt{\gamma_0}.
$

For a trial stepsize $\mu>0$, the projected SAPG point is
\begin{equation}
	\label{eq:sapg_sap_step}
	\mathbf x_k^{\mathrm{sap}}
	=
	(1-\eta)\mathbf x_k
	+
	\eta
	\Pi_{\mathcal X}
	\left(
	\mathbf x_k+\mu\mathbf p_k
	\right),
\end{equation}
where $0<\eta\le1$, and $\Pi_{\mathcal{X}}$ denotes the Euclidean projection map onto $\mathcal{X}$.
The stepsize is selected by  backtracking until
\begin{equation}
	\label{eq:sapg_acceptance}
	\left\{
	\begin{aligned}
	&\nabla F(\mathbf{x}_k)^T
(\mathbf x_k^{\mathrm{sap}}-\mathbf x_k)
>0,\\
		&F\bigl(\mathbf x_k^{\mathrm{sap}}\bigr)
\ge
F(\mathbf x_k)
+
c_{\mathrm A}
\nabla F(\mathbf{x}_k)^T
(\mathbf x_k^{\mathrm{sap}}-\mathbf x_k),
	\end{aligned}
	\right.
\end{equation}
where $0<c_{\mathrm A}<1$. The first condition verifies that the projected displacement $\mathbf x_k^{\mathrm{sap}}-\mathbf x_k$ retains the ascent property of $\mathbf p_k$, while the second is the Armijo sufficient-increase condition. 
To retain the standard projected-gradient convergence argument, SAPG includes a correction step when no projected SAPG point satisfies \eqref{eq:sapg_acceptance} within the prescribed number of backtracking reductions. The correction point is the conventional DPG mapping
\begin{equation}
	\label{eq:sapg_pg_step}
	\mathbf x_k^{\mathrm{pg}}
	=
	(1-\eta)\mathbf x_k
	+
	\eta
	\Pi_{\mathcal X}
	\left(
	\mathbf x_k+
	\mu\nabla F(\mathbf{x}_k)
	\right),
\end{equation}
and the local Lipschitz continuity of  $\nabla F$ gives
\begin{equation}
	\label{eq:sapg_pg_armijo}
	F\bigl(\mathbf x_k^{\mathrm{pg}} \bigr)
	\ge
	F(\mathbf x_k)
	+
	c_{\mathrm A}
	\nabla F(\mathbf{x}_k)^T
	(\mathbf x_k^{\mathrm{pg}}-\mathbf x_k)
\end{equation}
for all sufficiently small trial step size $\mu$. Specifically, if
$\mathbf x_k^{\mathrm{pg}}(\mu)-\mathbf x_k=\mathbf0$, then $\mathbf x_k$ is a fixed point of the projected-gradient map and satisfies the KKT condition of \eqref{eq:sapg_problem}.
The detailed algorithm is summarized in Algorithm~\ref{alg:sapg_nlos}.

%Thus, the soft-active metric determines the proposed search direction and incorporates the constraint information encoded by the weights in \eqref{eq:sapg_weights}. The projected-gradient correction is invoked only when Euclidean projection prevents the SAP displacement from satisfying the ascent conditions; under the local Lipschitz assumption, it guarantees either acceptance of a feasible ascent step or identification of a KKT point.

\begin{algorithm}[t]
	\caption{Proposed SAPG Algorithm}
	\label{alg:sapg_nlos}
	\renewcommand{\algorithmicrequire}{\textbf{Input:}}
	\renewcommand{\algorithmicensure}{\textbf{Output:}}
	\begin{algorithmic}[1]
		\footnotesize
		\REQUIRE Feasible APV $\mathbf x_0\in\mathcal X$; parameters
		$\eta$, $\gamma_0$, $\kappa$, $c_{\mathrm A}$, initial stepsize
		$\mu>0$, backtracking factor $\beta_{\mathrm{bt}}\in(0,1)$,
		$N_{\mathrm{sap}}$, $T_{\mathrm{SAP}}$, and $\epsilon_{\mathrm{pg}}$.
		\ENSURE APV $\mathbf x_\star$ and beamformer
		$\mathbf w_{\mathrm{opt}}(\mathbf x_\star)$.
		\STATE Set $k=0$.
		\REPEAT
		\STATE Evaluate $F(\mathbf x_k)$ and $\nabla F(\mathbf x_k)$
		by \eqref{eq:nu_star_eq_nlos_rev}-\eqref{eq:dFdx_basic_nlos_rev}.
		\IF{$k<T_{\mathrm{SAP}}$}
		\STATE Compute the slacks, $\tau_k$, and the weights by
		\eqref{eq:sapg_slacks}--\eqref{eq:sapg_weights}.
		\STATE Form $\mathbf H_k$ and solve
		$\mathbf H_k\mathbf p_k=\nabla F(\mathbf x_k)$.
		\STATE Apply at most $N_{\mathrm{sap}}$ backtracking steps
		to \eqref{eq:sapg_sap_step}.
		\ENDIF
		\IF{$k<T_{\mathrm{SAP}}$ and \eqref{eq:sapg_acceptance} holds}
		\STATE Set
		$\mathbf x_{k+1}
		=
		\mathbf x_k^{\mathrm{sap}}(\mu)$.
		\ELSE
		\STATE Apply backtracking to \eqref{eq:sapg_pg_step}
		until \eqref{eq:sapg_pg_armijo} holds.
		\STATE Set
		$\mathbf x_{k+1}
		=
		\mathbf x_k^{\mathrm{pg}}(\mu)$.
		\ENDIF
		\STATE Set $k\leftarrow k+1$.
		\UNTIL{$
			k>T_{\mathrm{SAP}}
			\ \text{and}\
			\dfrac{
				\|\mathbf x_k-\mathbf x_{k-1}\|_2
			}{
				\eta\mu\max\{1,\|\mathbf x_{k-1}\|_2\}
			}
			\le
			\epsilon_{\mathrm{pg}}
			$}
		\RETURN $\mathbf x_\star=\mathbf x_k$ and compute $\mathbf{w}_{\mathrm{opt}}(\mathbf{x}_{\star})$
		by \eqref{eq:wopt_nlos}. 
	\end{algorithmic}
\end{algorithm}

We now evaluate the computational complexity of
Algorithm~\ref{alg:sapg_nlos}. Evaluating the jammer and desired
channels costs $\mathcal O(N_r(I+L))$, forming
$\mathbf C(\mathbf x)$ and $\mathbf b(\mathbf x)$ costs
$\mathcal O(N_rI^2)$, and factorizing $\mathbf C(\mathbf x)$ by
Cholesky decomposition costs $\mathcal O(I^3)$. After solving
\eqref{eq:nu_star_eq_nlos_rev}, forming the residual and evaluating
its row-wise derivatives require $\mathcal O(N_r(I+L))$ operations.
Computing the slacks and weights, forming the tridiagonal matrix
$\mathbf H_k$, and solving
$\mathbf H_k\mathbf p_k=\nabla F(\mathbf x_k)$ each require
$\mathcal O(N_r)$ operations. The bounded PAVA projection also has
linear complexity. 
The final MVDR beamformer can be recovered without an
$N_r\times N_r$ matrix inversion. Specifically, combining
\eqref{eq:woodbury_nlos_rev},
\eqref{eq:nu_star_eq_nlos_rev}, and
\eqref{eq:g_equals_Jmin_nlos_rev} gives
\begin{equation}
	\label{eq:low_complexity_mvdr_recovery}
	\mathbf w_{\mathrm{opt}}(\mathbf x)
	=
	\frac{
		\mathbf h_0(\mathbf x)
		-
		\mathbf V(\mathbf x)\boldsymbol\nu^\star(\mathbf x)
	}{
		F(\mathbf x)
	}
	=
	-\frac{\mathbf r(\mathbf x)}{F(\mathbf x)}.
\end{equation}
Since $F(\mathbf x_\star)$ and $\mathbf r(\mathbf x_\star)$ are
already available from the final first-order evaluation, recovering
$\mathbf w_{\mathrm{opt}}(\mathbf x_\star)$ requires only
$\mathcal O(N_r)$ additional operations.
Consequently, the per-iteration complexity is
$\mathcal O(N_rI^2+I^3+N_r(I+L))$, and over $K$ iterations the
overall complexity of Algorithm~\ref{alg:sapg_nlos} is
$\mathcal O\!\left(
K[N_rI^2+I^3+N_r(I+L)]
\right)$.

\subsection{Extension to ZF Reception}
\label{subsec:mvdr_zf_unification}

The preceding antenna position optimization framework is not restricted to MVDR reception, but can also be applied to ZF reception, since the corresponding APV objective admits the same LS residual structure as in \eqref{eq:J_def_nlos_rev}.
To establish this connection,
first collect the jammer steering vectors as
\begin{equation}
	\label{eq:zf_jammer_matrix}
	\mathbf A_I(\mathbf x)
	\triangleq
	\left[
	\mathbf a(\mathbf x,\phi_1),
	\ldots,
	\mathbf a(\mathbf x,\phi_I)
	\right],
\end{equation}
and thus the ZF constraints
$\mathbf w^H\mathbf g_i(\mathbf x)=0$, $i=1,\ldots,I$,
are equivalent to
$
	\mathbf A_I^H(\mathbf x)\mathbf w=\mathbf 0.
$
Under this constraint, the optimal ZF
receiver for a fixed APV is therefore characterized by
\begin{equation}
	\label{eq:zf_receiver_problem}
	\begin{aligned}
		F_{\mathrm{ZF}}(\mathbf x)
		\triangleq
		\max_{\substack{\|\mathbf w\|_2=1,\\
				\mathbf A_I^H(\mathbf x)\mathbf w=\mathbf 0				
		}}
		|\mathbf w^H\mathbf h_0(\mathbf x)|^2,
	\end{aligned}
\end{equation}
which has a feasible solution when 
$\operatorname{rank}(\mathbf A_I(\mathbf x))<N_r$ and $\mathbf P_I^\perp(\mathbf x)\mathbf h_0(\mathbf x)
\neq\mathbf 0$, where $
\mathbf P_I^\perp(\mathbf x)
\triangleq
\mathbf I_{N_r}
-
\mathbf A_I(\mathbf x)
\mathbf A_I^\dagger(\mathbf x)
$ is the orthogonal projector onto the null space of
$\mathbf A_I^H(\mathbf x)$.

Under this condition, the solution of
\eqref{eq:zf_receiver_problem} is
\begin{equation}
	\label{eq:zf_beamformer}
	\mathbf w_{\mathrm{ZF}}(\mathbf x)
	=
	\frac{
		\mathbf P_I^\perp(\mathbf x)\mathbf h_0(\mathbf x)
	}{
		\|
		\mathbf P_I^\perp(\mathbf x)
		\mathbf h_0(\mathbf x)\|_2
	},
\end{equation}
and the resulting output SINR is
$
	\mathrm{SINR}_{\mathrm{ZF}}(\mathbf x)
	=
	\frac{\sigma_s^2}{\sigma_n^2}
	F_{\mathrm{ZF}}(\mathbf x),
$
where  $
	F_{\mathrm{ZF}}(\mathbf x)
	\triangleq
	\mathbf h_0^H(\mathbf x)
	\mathbf P_I^\perp(\mathbf x)
	\mathbf h_0(\mathbf x).
$
Hence, similar to \eqref{eq:P0_nlos}, the ZF-based APV optimization problem is
\begin{equation}
	\label{eq:zf_apv_problem}
	\max_{\mathbf x\in\mathcal X}
	F_{\mathrm{ZF}}(\mathbf x).
\end{equation}

The key observation is that
$F_{\mathrm{ZF}}(\mathbf x)$ is itself an LS residual.
Since
$\mathbf A_I(\mathbf x)\mathbf A_I^\dagger(\mathbf x)$
is the orthogonal projector onto the column space of
$\mathbf A_I(\mathbf x)$, the orthogonal projection theorem gives
\begin{equation}
	\label{eq:zf_residual_equivalence}
		F_{\mathrm{ZF}}(\mathbf x)
		=
		\left\|
		\mathbf P_I^\perp(\mathbf x)
		\mathbf h_0(\mathbf x)
		\right\|_2^2=
		\min_{\boldsymbol\nu\in\mathbb C^I}
		\left\|
		\mathbf A_I(\mathbf x)\boldsymbol\nu
		-
		\mathbf h_0(\mathbf x)
		\right\|_2^2.
\end{equation}
If $\mathbf A_I(\mathbf x)$ has full column rank throughout
$\mathcal X$, the inner LS problem in
\eqref{eq:zf_residual_equivalence} has the following unique solution:
\begin{equation}
	\label{eq:zf_ls_solution}
	\boldsymbol\nu_{\mathrm{ZF}}^\star(\mathbf x)
	=
	\left(
	\mathbf A_I^H(\mathbf x)\mathbf A_I(\mathbf x)
	\right)^{-1}
	\mathbf A_I^H(\mathbf x)\mathbf h_0(\mathbf x),
\end{equation}
which enables $\nabla F_{\mathrm{ZF}}(\mathbf x)$ to be obtained by the same Danskin
argument as in Theorem~\ref{thm:danskin_nlos_rev}. Consequently, the
ZF problem \eqref{eq:zf_apv_problem} can be handled by the same SAPG
framework.

During APV optimization, however, the columns of
$\mathbf A_I(\mathbf x)$ may become linearly dependent at some feasible array geometries, which may render the inner
LS minimizer in \eqref{eq:zf_residual_equivalence}
nonunique. Moreover, the pseudoinverse
$\mathbf A_I^\dagger(\mathbf x)$ need not vary smoothly with
$\mathbf x$ when the rank of $\mathbf A_I(\mathbf x)$ changes. To retain a unique inner solution
and a continuously differentiable value function, we introduce a regularized ZF objective as given by
\begin{equation}
	\label{eq:regularized_zf_objective}
	F_{\epsilon}^{\mathrm{ZF}}(\mathbf x)
	\triangleq
	\min_{\boldsymbol\nu\in\mathbb C^I}
	\left\{
	\left\|
	\mathbf A_I(\mathbf x)\boldsymbol\nu
	-
	\mathbf h_0(\mathbf x)
	\right\|_2^2
	+
	\epsilon\|\boldsymbol\nu\|_2^2
	\right\},
\end{equation}
which preserves the ridge LS structure of
\eqref{eq:J_def_nlos_rev}, with
$(\mathbf V(\mathbf x),\sigma_n^2)$ replaced by
$(\mathbf A_I(\mathbf x),\epsilon)$, where
$\epsilon>0$ is the regularization parameter.
The inner minimizer of \eqref{eq:regularized_zf_objective} is uniquely given by
\begin{equation}
	\label{eq:regularized_zf_solution}
	\boldsymbol\nu_{\epsilon}^\star(\mathbf x)
	=
	\left(
	\epsilon\mathbf I_I
	+
	\mathbf A_I^H(\mathbf x)\mathbf A_I(\mathbf x)
	\right)^{-1}
	\mathbf A_I^H(\mathbf x)\mathbf h_0(\mathbf x).
\end{equation}
Since the regularization term
is nonnegative, we have
\begin{equation}
	\label{eq:regularized_zf_error_bound}
	\begin{aligned}
		0
		&\le
		F_{\epsilon}^{\mathrm{ZF}}(\mathbf x)
		-
		F_{\mathrm{ZF}}(\mathbf x)\le
		\epsilon
		\left\|
		\mathbf A_I^\dagger(\mathbf x)
		\mathbf h_0(\mathbf x)
		\right\|_2^2.
	\end{aligned}
\end{equation}
Consequently, for every fixed APV,
\begin{equation}
	\label{eq:regularized_zf_limit}
	\lim_{\epsilon\downarrow0}
	F_{\epsilon}^{\mathrm{ZF}}(\mathbf x)
	=
	F_{\mathrm{ZF}}(\mathbf x),
\end{equation}
which shows that the proposed
first-order optimization framework extends naturally to ZF
reception. The resulting computational complexity remains of the
same order as that in the previous subsection.

\vspace{-0.2cm}
\section{Numerical Results}

This section evaluates the performance of the proposed MA-enabled anti-jamming design.
Unless otherwise specified, we set $N_r=8$, $\lambda=1$, $d=\lambda/2$,
$D_x=12\lambda$, $\sigma_n^2=1$, $\mathrm{SNR}=0$ dB, and
$\mathrm{JNR}=20$ dB. The desired-path AoAs
$\{\theta_\ell\}_{\ell=1}^{L}$ and the jammer AoAs
$\{\phi_i\}_{i=1}^{I}$ are independently drawn from the uniform
distribution over $[-\pi/2,\pi/2)$. The corresponding channel gains are
independently generated according to
$\alpha_\ell\sim\mathcal{CN}(0,1/L)$ and
$\zeta_i\sim\mathcal{CN}(0,1)$. Each reported point is averaged over
$10^4$ independent Monte Carlo realizations.
For the LoS single-jammer scenario, we set $L=I=1$. For DPGD, the
Lipschitz constant and algorithm parameters are chosen as
$L_a=4N_r a^2$, $\mu=0.1/L_a$, $\eta=\min\{0.9,0.99/(1+2\mu L_a)\}$, and the stopping tolerance is $10^{-7}$.
We compare the
proposed DPGD algorithm with the following four benchmark schemes:
1) the half-wavelength-spaced uniform linear array, denoted by ULAH,
with $x_n=(n-1)\lambda/2$;
2) the full-aperture uniform linear array, denoted by ULAF, with
$x_n=(n-1)D_x/(N_r-1)$;
3) the AO method, which alternates between
the closed-form MVDR beamformer update and the APV subproblem of the
original formulation in \eqref{eq:8}, which serves as a high-quality but high-complexity benchmark adopted in
\cite{TVTantiJamming}; and
4) the IPM that is directly applied to the reduced
problem in \eqref{eq:reduced_problem}.
For the general multipath multi-jammer scenario, we set $L=6$ and
$I=4$. The SAPG algorithm parameters are chosen as
 $\eta=0.8$, $T_{\mathrm{SAP}}=3000$, $\gamma_0=100,
\kappa=10^{-3},
c_{\mathrm A}=10^{-4},
\beta_{\mathrm{bt}}=0.5.$
We compare the proposed SAPG algorithm with five benchmark schemes:
1) damped projected-gradient ascent (DPGA), which replaces the proposed soft-active metric matrix $\mathbf H_k$ with the
identity matrix and therefore employs the conventional Euclidean
projected-gradient mapping in \eqref{eq:sapg_pg_step};
2) ULAH;
3) ULAF;
4) AO; and
5) IPM.

\begin{figure}[t]
	\centering
	\includegraphics[width=3.2in]{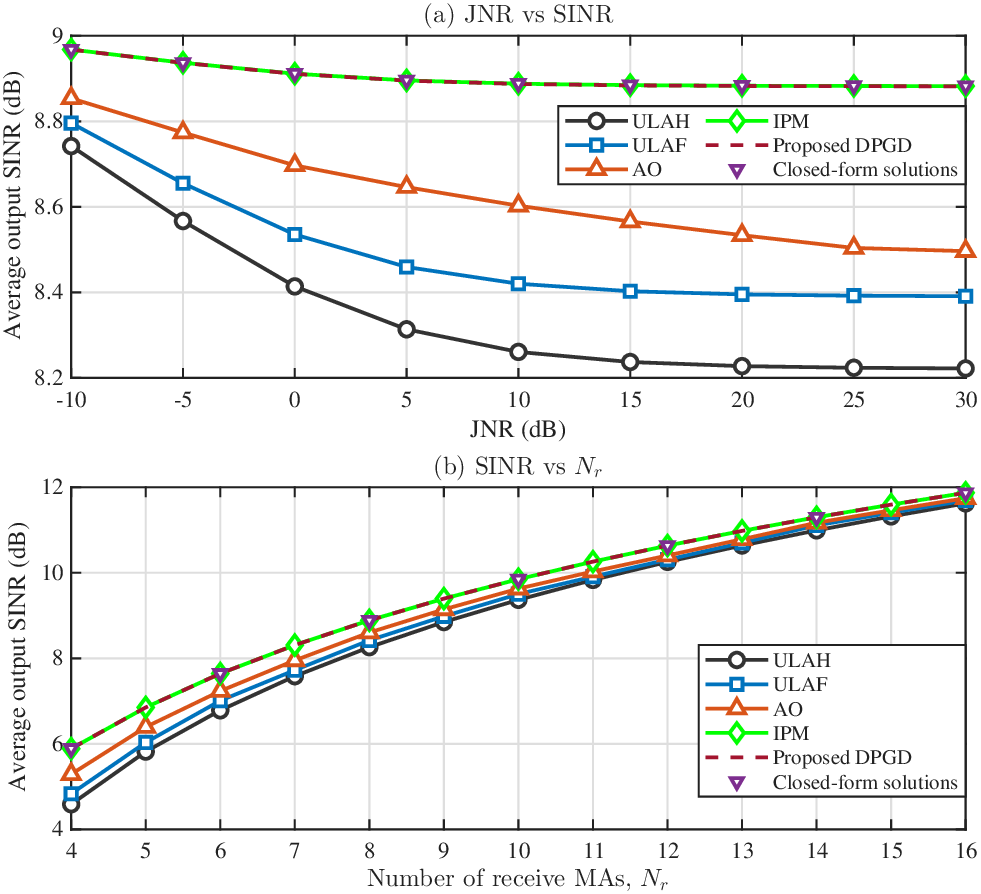}
	\caption{Average output SINR versus JNR and the number of receive MAs.}
	\label{fig:sinr_results}\vspace{-0.2cm}
\end{figure}

\begin{figure*}[!t]
	\centering
	\includegraphics[width=0.95\textwidth]{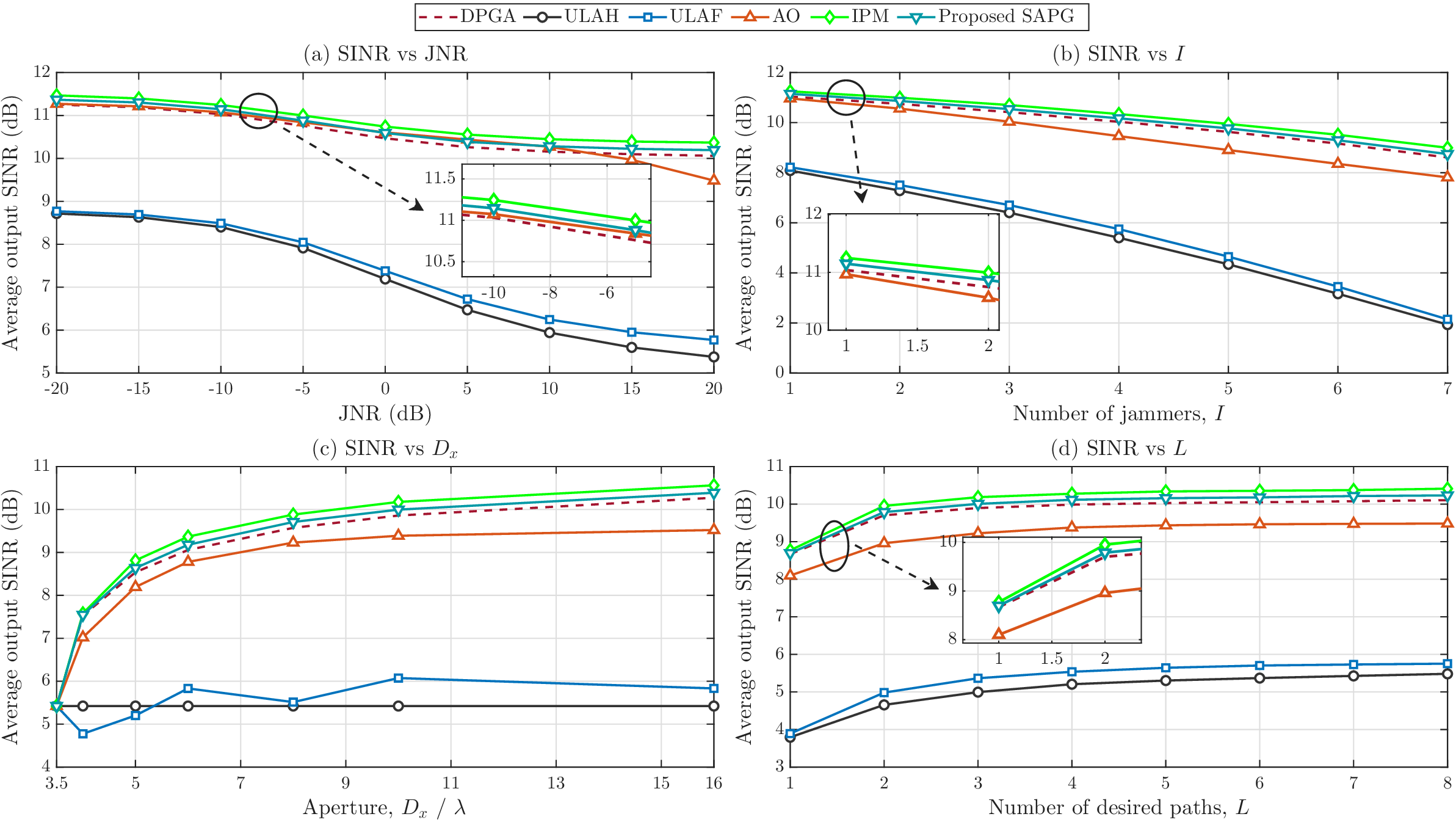}
	\caption{Average output SINR comparison in the general multipath multi-jammer scenario versus (a) the JNR, (b) the number of jammers $I$, (c) the normalized aperture $D_x/\lambda$, and (d) the number of desired paths $L$.}
	\label{fig:nlos_sinr_results}\vspace{-0.2cm}
\end{figure*}

%\begin{table}[t]
%	\caption{Average runtime versus the number of receive MAs (seconds).}
%	\label{tab:runtime_vs_Nr}
%	\centering
%	\begin{tabular}{c|ccc}
%		\hline
%		$N_r$ & AO & IPM & Proposed DPGD \\
%		\hlineS
%		4  & 0.0241 & 0.0130 & 0.0062 \\
%		6  & 0.0391 & 0.0389 & 0.0049 \\
%		8  & 0.0620 & 0.0437 & 0.0048 \\
%		10 & 0.1135 & 0.0482 & 0.0050 \\
%		12 & 0.1271 & 0.0532 & 0.0051 \\
%		\hline
%	\end{tabular}
%\end{table}

 \begin{table}[t]
	\caption{Average runtime versus the number of receive MAs under LoS channel (seconds). The aperture is set to $D_x=32\lambda$.}
	\label{tab:runtime_los_vs_Nr}
	\centering
	\begin{tabular}{c|ccc}
		\hline
		$N_r$ & AO & IPM & Proposed DPGD \\
		\hline
		4  & 0.0283 & 0.0156 & 0.0047 \\
		6  & 0.0446 & 0.0451 & 0.0052 \\
		8  & 0.0548 & 0.0540 & 0.0051 \\
		10 & 0.0605 & 0.0559 & 0.0052 \\
		12 & 0.0646 & 0.0575 & 0.0054 \\
		16 & 0.0782 & 0.0608 & 0.0059 \\
		25 & 0.1304 & 0.0799 & 0.0064 \\
		32 & 0.1915 & 0.0927 & 0.0074 \\
		\hline
	\end{tabular}\vspace{-0.2cm}
\end{table}

\begin{table}[t]
	\caption{Average runtime versus the number of receive MAs under multipath channel (seconds). The aperture is set to $D_x=32\lambda$.}
	\label{tab:runtime_nlos_vs_Nr}
	\centering
	\begin{tabular}{c|cccc}
		\hline
		$N_r$ & AO & IPM & DPGA & Proposed SAPG \\
		\hline
		4  & 0.2913 & 0.0177 & 0.0184 & 0.0069 \\
		6  & 0.4439 & 0.0296 & 0.0193 & 0.0117 \\
		8  & 0.6178 & 0.0428 & 0.0202 & 0.0163 \\
		10 & 0.8318 & 0.0682 & 0.0208 & 0.0175 \\
		12 & 1.0740 & 0.0931 & 0.0216 & 0.0212 \\
		16 & 1.7226 & 0.1652 & 0.0233 & 0.0223 \\
		25 & 4.5965 & 0.4836 & 0.0278 & 0.0238 \\
		32 & 6.3292 & 0.7355 & 0.0330 & 0.0242 \\
		\hline
	\end{tabular}\vspace{-0.2cm}
\end{table}

For the canonical setting with a single LoS path on the legitimate link and one jammer, Fig.~\ref{fig:sinr_results} (a) compares the average output SINR versus the JNR, while Fig.~\ref{fig:sinr_results} (b) evaluates the impact of the number of receive MAs, \(N_r\).
As shown in Fig.~\ref{fig:sinr_results} (a), the two fixed ULA baselines degrade as the JNR increases, since their array geometries cannot adapt to the jammer direction. In contrast, the MA-optimized schemes maintain higher SINR by reducing the steering coherence between the desired and jamming channels. Fig.~\ref{fig:sinr_results} (b) further shows that increasing $N_r$ improves the spatial degrees of freedom, and the optimized MA schemes consistently outperform the FPA baselines.
{{To further assess the solution quality of DPGD, we also include
	the available closed-form solutions as a benchmark. Since these
	solutions apply only to specific parameter regimes, the values of
	$N_r$ and $D_x$ used in this comparison are deliberately chosen
	to satisfy the corresponding closed-form conditions in
	\cite{ZeroForcing,AF_MA}. As observed in both
	Fig.~\ref{fig:sinr_results} (a) and (b), the proposed DPGD
	completely overlaps with the closed-form benchmark. Hence, although
	Theorem~\ref{thm:dpgd_convergence} guarantees only almost-sure
	convergence to a local minimizer, DPGD also exhibits globally optimal performance in these tested cases.}}
Compared with the other MA-based schemes, the inferior performance
of AO is mainly due to its alternating updates, in which the APV is
optimized for the current receive beamformer at each iteration, so
the algorithm generally converges only to a coordinate-wise
stationary point.
Nevertheless, while the output SINR
increases with $N_r$, the performance gaps among the compared schemes gradually narrow.
This trend can be explained by two
complementary effects. First, with fixed $D_x$ and $d$, the residual aperture available for position adaptation is
$D_x-(N_r-1)d$, which decreases as more antennas are accommodated within the same aperture.
Second, a single jammer spans at
most a one-dimensional interference subspace, and consequently, the additional benefit obtainable from APV adaptation by MA
becomes smaller. This observation is consistent with
\eqref{eq:mvdr_sinr_closed_form}, where MA repositioning does not
increase the nominal array-gain term $N_r$ but only reduces the
rank-one steering-coherence penalty.
Furthermore, in both Fig.~\ref{fig:sinr_results} (a) and (b), the performance of the proposed DPGD method overlaps with that of IPM. 
The close performance between DPGD and IPM provides further numerical support for the effectiveness of DPGD. Since IPM exploits second-order curvature information, it can use local curvature when navigating saddle-type regions, whereas DPGD relies only on first-order information. The comparable output SINR achieved by DPGD is therefore consistent with its almost-sure convergence to a local minimizer established in Theorem~\ref{thm:dpgd_convergence}.
Table~\ref{tab:runtime_los_vs_Nr} compares the average runtime of the optimization methods as the number of receive MAs increases. DPGD consistently requires the shortest runtime, and its computational advantage becomes more pronounced for larger arrays. In particular, at \(N_r=32\), DPGD is approximately 26 times and 13 times faster than AO and IPM, respectively.
Overall, these results support the preceding analysis in Section III and demonstrate that DPGD achieves high anti-jamming performance with substantially improved computational efficiency.

For the general setting allowing multiple propagation paths on the legitimate link and multiple jammers, Figs.~\ref{fig:nlos_sinr_results} (a) and (b) show the average output SINR versus the JNR and the number of jammers, respectively. In both figures, SAPG remains close to IPM and consistently outperforms the fixed-array, AO, and DPGA baselines. 
In Fig.~\ref{fig:nlos_sinr_results} (a), the SINR curves gradually flatten at high JNRs. When the jammer power is sufficiently large, the MVDR receiver places strong suppression on the jammer subspace. As a result, the desired-channel component overlapping with this subspace contributes little to the receiver output, and the remaining SINR is mainly determined by the desired-channel component outside the jammer subspace. Further increasing the jammer power then has only a limited effect on the output SINR. 
Fig.~\ref{fig:nlos_sinr_results} (b) shows a different effect when the number of jammers increases.
Additional jammer directions generally enlarge the jammer subspace, leaving less spatial freedom to preserve the desired signal while suppressing interference.
This also explains why the performance gap
between the optimized MA schemes and the fixed arrays becomes more
pronounced in the multi-jammer regime.
By adjusting the antenna positions, the MA receiver can reduce the overlap between the desired channel and the enlarged jammer subspace. The performance advantage of the optimized MA schemes therefore becomes more evident as more jammers are present.

Figs.~\ref{fig:nlos_sinr_results} (c) and (d) show the average output SINR versus the normalized aperture $D_x/\lambda$ and the number of desired paths $L$, respectively. SAPG maintains performance comparable to IPM and consistently outperforms the other baselines over the tested ranges. As shown in Fig.~\ref{fig:nlos_sinr_results} (c), enlarging the aperture improves the output SINR because the MAs have greater freedom to adjust the array geometry, although the gain gradually saturates as the aperture becomes sufficiently large. Fig.~\ref{fig:nlos_sinr_results} (d) shows a similar trend as $L$ increases. Since the path gains are normalized, the improvement mainly comes from the richer multipath phase structure that can be exploited through APV adaptation rather than from an increase in the total desired-channel power. Table~\ref{tab:runtime_nlos_vs_Nr} compares the average runtime of the optimization methods as the number of receive MAs increases. SAPG requires the shortest runtime for all tested array sizes, with its computational advantage becoming more pronounced as $N_r$ increases. At $N_r=32$, SAPG is approximately 262 times and 31 times faster than AO and IPM, respectively.
Overall, these results validate SAPG in the general setting, showing strong anti-jamming performance and substantial computational efficiency gains under various channel and interference conditions.

\vspace{-0.2cm}
\section{Conclusion}

This paper developed low-complexity optimization methods for
MA-enabled anti-jamming. For the LoS single-jammer case, the
joint MVDR beamformer and APV design was reduced exactly to a
steering-coherence minimization problem. Based on its generic active
strict-saddle landscape, we developed a linear-complexity DPGD
algorithm and proved its almost-sure convergence to a local
minimizer. For the general multipath multi-jammer case, an equivalent  
ridge LS reformulation was proposed to enable efficient gradient
evaluation and lead to the proposed SAPG algorithm, while the same
residual structure further extended the optimization framework to the case with
ZF reception. Numerical results demonstrated that the proposed
algorithms attain IPM-level anti-jamming performance with
substantially lower computational cost, while consistently outperforming the considered fixed-array and competing optimization schemes.
%{Extending the proposed antenna position optimization approach to two-dimensional arrays and 
%multiuser/multistream reception \cite{MultiuserCommunicationsMA}, where the APV jointly affects
%multiple desired signals, remains a possible direction for future work.}
Extending the proposed antenna position optimization approach to
two-dimensional arrays and to multiuser/multistream reception,
as considered in related MA communication settings
\cite{MultiuserCommunicationsMA}, is an interesting direction for
future work.

\vspace{-0.2cm}
\bibliography{ref1.bib}
\bibliographystyle{IEEEtran}

\end{document}